\documentclass[11pt,preprint]{aastex}

\def\'#1{\ifx#1i{\accent"13\i}\else{\accent"13#1}\fi}

\def\alamenos#1{$^{-#1}$}

\newcommand\msun{\rm M_{\odot}}

\newcommand\be{\begin{equation}}
\newcommand\en{\end{equation}}
\newcommand\cm{\rm cm}
\newcommand\kms{\rm{\, km \, s^{-1}}}
\newcommand\K{\rm K}
\def\h2{\rm H_2}
\newcommand\etal{{\rm et al}.\ }

\begin{document}

\title{Rapid Formation of Molecular Clouds and Stars in the Solar
Neighborhood} 

\author{Lee Hartmann\altaffilmark{1}, Javier
Ballesteros-Paredes\altaffilmark{2,3}, and Edwin
A. Bergin\altaffilmark{1}}  
\altaffiltext{1}{Harvard-Smithsonian Center for Astrophysics, 60
Garden St., Cambridge, MA 02138;\\ Electronic mail:
hartmann@cfa.harvard.edu, ebergin@cfa.harvard.edu} 
\altaffiltext{2}{American Museum of Natural History, 79th St. at
Central Park W., New York, NY 10024; javierbp@amnh.org} 
\altaffiltext{3}{Instituto de Astronom\'{\i}a, UNAM, Ap. Postal
70-264, Cd. Universitaria, 04510 M\'exico D.F., M\'exico} 

\begin{abstract}
We show how molecular clouds in the solar neighborhood
might be formed and produce stars rapidly enough to explain 
stellar population ages,
building on results from numerical simulations of the turbulent interstellar 
medium and general considerations of molecular gas formation. 
Observations of both star-forming regions and young, gas-free stellar
associations indicate that most nearby molecular clouds form stars
only over a short time span before dispersal; large-scale flows in the
diffuse interstellar medium have the potential for forming clouds
sufficiently rapidly, and for producing stellar populations with ages much less
than the lateral crossing times of their host molecular clouds. We
identify four important factors for understanding rapid star formation and
short cloud lifetimes. First, much of the accumulation and dispersal
of clouds near the solar circle might occur in the atomic phase; only
the high-density portion of a cloud's lifecycle is spent in the
molecular phase, thus helping to limit molecular cloud
``lifetimes''. Second, once a cloud achieves a high enough column
density to form $\h2$ and CO, gravitational forces become larger than
typical interstellar pressure forces; thus star formation can follow rapidly upon
molecular gas formation and turbulent dissipation in limited areas of
each cloud complex. Third, typical magnetic fields are not strong
enough to prevent rapid cloud formation and gravitational
collapse. Fourth, rapid dispersal of gas by newly-formed stars,
passing shock waves, and reduction of shielding by a small expansion
of the cloud after the first events of star formation, might limit the
length of the star formation epoch and the lifetime of a cloud in its
molecular state.  This picture emphasizes the importance of
large-scale boundary conditions for understanding molecular cloud
formation; implies that star formation is a highly dynamic, rather
than quasi-static, process; and that the low galactic star formation
rate is due to low efficiency rather than slowed collapse in local regions. 
\end{abstract}


\section{Introduction}

This paper has its origin in the long effort to answer the following
deceptively simple question: where are the ``post-T Tauri stars'' in
the Taurus molecular clouds?  As originally suggested by Herbig
(1978), there was every reason to expect that the stellar population
of Taurus would exhibit a substantial spread in ages, especially if
molecular cloud complexes last for several tens of Myr (Elmegreen
1991), and if ambipolar diffusion of magnetic flux over 5-10 Myr is
necessary before molecular cloud cores can collapse to form stars
(Mouschovias 1991; Shu, Adams, \& Lizano 1987). Yet several surveys,
beginning with Herbig, Vrba, \& Rydgren (1986) found no evidence for a
substantial population of post-T Tauri stars (PTTS) with ages $>
3$~Myr (Hartmann \etal 1991; Gomez \etal 1992; Brice\~no \etal 1997,
1999).  X-ray surveys of the area did find older stars (Walter \etal
1988; Neuh\"auser \etal 1995; Wichmann \etal 1996), but these stars
are too dispersed spatially, and are too old ($\sim 10-100$~Myr), to
constitute the ``missing'' $>$ 5 Myr-old PTTS of the Taurus clouds
(Brice\~no \etal 1997; Feigelson 1996; see \S 2.1). 

Taurus is not exceptional.  Most star-forming regions containing
molecular gas have populations with typical ages $\sim 1-3$ Myr, and
very few (if any) stars of greater age (see the recent analysis by
Palla \& Stahler 2000; also Hartmann 2001). Moreover, of the
substantial molecular cloud complexes within about 350 pc -- Taurus,
Ophiuchus, Chamaeleon, Corona Australis, Lupus, Serpens, Perseus --
only one, the Coalsack, exhibits little or no evidence for young stars
(Nyman 1991; but see Eaton \etal 1990); this indicates that star
formation follows the formation of a typical molecular cloud complex
within less than 1 Myr (\S 2.1). The age dispersion of stars in
clusters suggests that molecular cloud complexes have lifetimes of the
order of their dynamical timescales (Elmegreen 2000). Moreover, the
absence of $\sim 5-10$~Myr-old stars in star-forming regions suggests
that molecular cloud complexes in the solar neighborhood must coalesce
rapidly, form stars rapidly, and disperse rapidly (Hartmann 2000). The
view that molecular clouds are relatively transient features goes back
at least to the seminal paper by Larson (1981); advances in
characterizing the young stellar populations of molecular clouds over
the last 20 years reinforce this conclusion.

Rapid cloud evolution has important consequences for understanding the
physics of molecular clouds and star formation. If cloud lifetimes are
short, there is no need for maintaining a quasi-equilibrium in
molecular clouds, so that MHD turbulence (which decays rapidly; Stone,
Ostriker, \& Gammie 1998; Mac Low \etal 1998; Mac Low 1999) need not
be regenerated (Ballesteros-Paredes, Hartmann \&\ V\'azquez-Semadeni
1999 = BHV; Elmegreen 2000).  Indeed, the limitations placed on cloud
lifetimes by stellar populations essentially ensure that the evolution
from dispersed molecular gas to protostellar cores is dynamic rather
than quasi-static (Ballesteros-Paredes, V\'azquez-Semadeni \&\ Scalo
1999 = BVS; Klessen, Heitsch, \& Mac Low 2000; Padoan \etal 2001). The
low galactic rate of star formation must be the result of reduced
efficiency in conversion of gas to stars (Hartmann 1998; Elmegreen
2000) rather than slowing collapse by strong magnetic support.

Solving the post-T Tauri problem by making cloud lifetimes short,
however, raises a new set of questions: how are molecular clouds
formed so rapidly, and why is it that the clouds form stars so
readily, especially given the potential restraining effects of
magnetic fields?  We argue that
formation of clouds and triggering of star formation by large-scale
flows, as we argued previously for Taurus (BHV; see also Sasao 1973;
Elmegreen 1993; Scalo \& Chappell 1999) is essential to
forming clouds and stars on less than a lateral crossing time.  
We further show that the conditions needed for molecular gas formation from
atomic flows -- a minimum column density for shielding, and limited
turbulent and magnetic support to achieve high enough densities for
reasonably rapid chemical evolution -- are similar to the conditions
needed for gravitational instability with collapse times of order a
Myr; therefore, star formation can commence rapidly once molecular
clouds are produced.  We also use general arguments and the results of
numerical simulations of the interstellar medium to show that
molecular clouds are probably supercritical, and thus magnetic fields
do not significantly slow gravitational collapse. 

The organization of this paper is as follows.  In \S 2 we summarize 
observational constraints from stellar population ages on cloud and star
formation timescales. In \S 3 we outline a physical scenario in agreement with
observations, presenting some numerical results which help support
this picture. We consider some further implications of rapid cloud and
star formation in \S 4; and in \S 5 we summarize our conclusions.

\section{Stellar population constraints}

\subsection{Stellar ages in associations}

The absence of many stars in Taurus and other star forming regions
older than about 3-5 Myr was discussed in detail in BHV.  In view of
the importance of the result, and the implications of this
best-studied cloud for the interpretation of observations of more
distant regions, we revisit this issue.  We also incorporate updated
pre-main sequence stellar isochrones.  As discussed, for example, in
Hartmann (2001), the principal systematic error in ages for most T
Tauri stars is uncertainty in the stellar mass.  Recent recent
measurements of T Tauri masses using disk rotation (Simon, Dutrey, \&
Guilloteau 2000) suggest that earlier isochrones underestimated the
stellar masses and ages; the net effect of the new estimates, which
we use here, is to make the low-mass stars approximately a factor of 
two older than one would infer using the D'Antona \& Mazzitelli (1994) 
tracks as in BHV. 

We begin by focussing on the issue of X-ray detected stars in
Taurus. Neuh\"auser \etal (1995) and Wichmann \etal (1996) identified
a number of potential pre-main sequence stars in the general direction
of Taurus from the ROSAT All-Sky Survey (RASS), and suggested that
these objects represent a significant, older population of
Taurus. However, Brice\~no \etal (1997) pointed out that the mere
presence of Li absorption was not a certain indicator of pre-main
sequence stars, since Li is not strongly depleted in G and early-K
stars until ages greater than $\sim 100$~Myr.  Moreover, Brice\~no
\etal showed that ROSAT was almost equally sensitive to 100 Myr-old
stars as to 1-10 Myr-old stars, thus the ROSAT survey must include
many stars much older than 10 Myr.  In addition, the observed number of
ROSAT stellar sources agrees with the number expected for an average
formation rate in the solar neighborhood over a 100 Myr period.
Brice\~no \etal thus argued that most of the ROSAT sources are much
older than Taurus and thus originated in different clouds which no
longer exist.  These stars then would disperse considerably over many
tens of Myr, thus explaining their smooth spatial distribution and
lack of concentration near Taurus.  This point of view was supported
by Mart\'in \& Magazzu (1999), who conducted a careful analysis of Li
equivalent widths and concluded that only about 22\% of the RASS stars
are probably pre-main sequence stars.  

In a followup detailed study, Wichmann \etal (2000) examined 58 of the
72 RASS weak-emission stars with Li found near Taurus, and
argued that approximately 60\% of these were true pre-main sequence
objects, somewhat higher than the proportion estimated by Mart\'in \&
Magazzu (1999).  However, it is important to place these objects in
the appropriate context to understand their significance.  To do this
consider Fig.~\ref{fig:HR}, where we place
these objects in the $V$ vs. $V-I$ color-magnitude diagram for that
portion of the sample (about half) for which Wichmann \etal (2000)
present photometry. We also plot isochrones from Siess, Dufour, \&
Forestini (2000) which are in better agreement with the recent T Tauri
mass measurements from disk rotation, and agree fairly well with the
Palla \& Stahler (2000) results. 

The most important result to note from Figure 2 is that most of the
RASS pre-main sequence (PMS) stars are much older than 10 Myr, {\em
assuming that they are at the distance of Taurus.}  Therefore they
cannot explain the drop in star formation rate beyond 3-4 Myr found by
Kenyon \& Hartmann (1990), BHV, Palla \& Stahler (2000), and Hartmann
(2001).  Indeed, because there are significantly fewer RASS stars than
T Tauri stars in the same mass range, these objects cannot represent a
significant local star formation rate.  At a constant rate of star
formation, there should be $\sim 10$ times as many 30 Myr-old stars as
3 Myr-old stars, and this is far from the case, even considering that
only about half of the possible RASS sources have photometry and thus
are plotted in Figure 2 (a substantial number of Taurus TTS also lack
$V-I$ colors and thus are missing as well).  This essential point
concerning the star formation rate in Taurus and many other regions
has been emphasized strongly by Palla \& Stahler (2000). 
 
One should also note that the ages suggested by placement of RASS
sources in the color-magnitude diagram are likely to be {\em lower
limits}.  The isochrones assume a distance of 140 pc, and any
foreground stars will appear to be too young (the RASS is biased
toward foreground stars; Brice\~no \etal 1997). This possibility is
further strengthened by considering the RASS zero-age main sequence
sources (ZAMS), defined by Wichmann \etal (2000) based on their Li
equivalent widths. These stars (crosses in Figure 2) generally lie
well above the main sequence for a distance of 140 pc because they are actually much
closer than 140 pc.  It would be surprising if the high-Li star
sample, selected by the same X-ray criteria, did not have the same
bias. 

Therefore, we must conclude that the majority of the RASS sources do
not belong to the present Taurus molecular cloud complex, but must
have originated in a much larger area from clouds that no longer
exist, as for example those which constituted the Gould belt (Wichmann
\etal 2000). 

While Taurus is by far the best-studied star-forming region for the
purpose of searching for older populations, studies of all other
nearby regions yield the same result.  Table 1 is a compilation of results
from the literature, demonstrating that stellar populations lying
within molecular clouds have average ages no larger than about 3 Myr;
the molecular gas has been dispersed in some way from older regions.
While one can find small numbers of ``older'' stars in all these
regions, the results of Palla \& Stahler (2000) show that these do not
imply significant star formation rates.  Moreover, as emphasized by
Hartmann (2001), the errors in placing stars in the HR or
color-magnitude diagrams are sufficiently large that one should expect
to find a few objects with spuriously large ages.  Finally, as
illustrated in Figure 2, it is extremely easy to contaminate a
pre-main sequence sample with older foreground populations that 
did not originate from presently-existing molecular gas.

All these considerations reinforce the inference of rapid star
formation over limited age spans in molecular clouds.  Furthermore,
the lack of substantial aggregations of molecular gas without star
formation (the Coalsack seems to be the only local exception)
indicates that star formation must ensue rapidly upon cloud formation.
These time limits place severe constraints on understanding cloud and
star formation. 

\subsection{The crossing time problem and large-scale flows}

The observational result that poses the greatest challenge to theory
is that both the inferred delay time between cloud formation and star
formation and the ages of the young stars present can be considerably
smaller than the lateral crossing time or dynamical time of the
star-formation region, suggesting that some kind of external
``triggering'' must be involved. 

For instance, consider the Taurus molecular cloud complex. The
projected extent of the Taurus clouds is between $\sim 20$ and 40 pc
(Figure 1), depending upon whether all the outlying regions of
molecular gas are considered to be part of the complex.  With a
typical velocity dispersion of $\sim 2$~$\kms$, as determined from
either the molecular gas motions (e.g., Kleiner \& Dickman 1985), or
from the velocity dispersion of the stellar population (Jones \&
Herbig 1979), the lateral crossing time of the Taurus complex is then
10-20 Myr. Yet, using the most up-to-date calibrations of stellar
evolutionary tracks (e.g., Simon, Dutrey, \& Guilloteau 2000), the
average age of the stars in Taurus is only about 2 Myr, and the ages
of the vast majority of the stars are less than $\sim$~4 Myr
(Palla \& Stahler 2000; Hartmann 2001; White \& Ghez 2001). 

While the youth of stars in molecular clouds has been recognized as a
problem for standard theories of low-mass star formation (e.g.,
Hartmann \etal 1991; Feigelson 1996; Brice\~no \etal 1997; Palla \&
Galli 1997), the constraints posed by gas-free but relatively youthful
stellar associations have been underappreciated. Consider the Sco OB2
association, which consists of three subgroups spanning a total length
of $\sim$ 150 pc (Blaauw 1960, 1964; Blaauw 1991; de Zeeuw \etal
1999; de Bruijn 1999). Assuming that the (one-dimensional) stellar
velocity dispersion of only $\lesssim 1.5$~$\kms$ found by de Bruijne
(1999) from an analysis of Hipparcos data is representative of the
motions in the parent molecular cloud which formed Sco OB2 (as in
Taurus; see above paragraph), the lateral crossing time of the entire
complex is $\sim 150$ Myr. \footnote{Mamajek, Lawson, \& Feigelson
(2000) suggest that the Sco OB2 subgroups may have been somewhat
closer together ($\sim 100$~pc span) about 10 Myr ago, but de Bruijn
(personal communication) finds that the proper motions of the subgroups are
the same within the observational errors.  In any case this reduction
in the initial size of the association still requires a propagation
speed for any external trigger of 7-10~$\kms$ along the long
dimension, still much larger than the observed stellar velocity
dispersion.} However, the ages of the Sco OB2 subgroups span a range
of only 10 Myr (de Geus \etal 1989; Preibisch \& Zinnecker 1999;
Preibisch, G\"unther, \& Zinnecker 2001), or about 15 Myr if one
includes the $\sim 1$~Myr-old Ophiuchus molecular cloud complex at one
end of the association.

The obvious conclusion to be drawn is that the lateral crossing times
of elongated star-forming regions are (often) irrelevant to formation
processes. What is required is some mechanism of triggering the onset
of cloud and star formation which operates externally, spans large
scales, and does not require the propagation of information laterally
to trigger star formation. 

Recently some of us (BHV) 
presented observational evidence and theoretical arguments suggesting
that the Taurus molecular cloud complex (and by inference, many other
star-forming regions) could have been formed very rapidly by
converging supersonic {\em large-scale} flows powered by the global
energy input from previous episodes of star formation.  The importance
of large-scale turbulence and flows has previously been emphasized by
Sasao (1973), Elmegreen (1993), and Scalo \& Chappell (1999).  In this
picture, molecular clouds are formed in the post-shock gas with a
lateral extent set by the coherence in the large scale velocity field;
then the relevant crossing or dynamical time is that of the shortest
dimension, not the longest.

While the idea that cloud and star formation can be triggered by flows
driven by massive stars is hardly new (e.g., Blaauw 1964, 1991;
Elmegreen \& Lada 1977; McCray \& Kafatos 1987), this picture (see
also V\'azquez-Semadeni, Passot \&\ Pouquet 1995 = VPP) differs from
some other scenarios by recognizing that stellar energy input can
drive flows over very large scales, i.e.\ hundreds of pc, as in the
``supershell'' picture of McCray \& Kafatos (1987).
 In this picture, the global stellar energy input feeds the turbulence
at small scales, but local bubbles expand and interact with their
surrounding medium such that the morphology of the structures can
become complicated (and in general do not exhibit simple bubble geometry). 

The large-scale nature of the flow field has several important
consequences. Large-scale motions make it much easier to form
extremely large structures in which star formation can be triggered
nearly simultaneously. The simplest example of this is a shell driven
outwards by winds and supernovae from massive stars in a central
cluster (Figure 3, left panel). The expanding shell eventually sweeps
up enough mass to become gravitationally unstable and form stars.  If
the radius of the shell or bubble is large enough, star formation can
be coordinated over very large distances, as a result of having a
common event driving evolution in the radial direction, rather than
any propagation of compression along the shell.  An example of this
type of structure may be found in the Cep OB2 association (Patel \etal
1995, 1998). 

A second consequence of the large-scale nature of the motions in the
interstellar medium is that flows driven by different regions interact
to form complex structures that in general are not simply related to
specific triggering sites (Figure 3, right panel). Thus, the Taurus
complex may have been triggered by interacting flows even though it
looks nothing like a ring or shell (Figure~1). 

In contrast to triggering by non-local flows, local triggering models
appear to have some difficulties in explaining spatially-extended
associations with small velocity dispersions like Taurus and Sco OB2.
Propagation speeds of 10-15~$\kms$ are needed to trigger star
formation across these regions within the required age spans. It is
clear that the molecular gas which is forming stars in Taurus, or has
formed stars in Sco OB2, cannot have been moving this fast; there is
no evidence for such velocity dispersions and gradients from the
stellar proper motions in Taurus and Sco OB2 (Jones \& Herbig 1979; de
Bruijn 1999).  Therefore, the models of Blaauw (1964, 1991), Elmegreen
\& Lada (1977), de Geus \etal (1989), and Prebisch \& Zinnecker (1999)
for Sco OB2 require that the large scale molecular cloud must have
existed long before the local triggering by stellar energy input
occurred.  In other words, in local triggering models the cloud {\em
formation} process is independent of the {\em star formation} process.
But these models further demand no significant star formation in the
extended molecular cloud, over a period of 5-10 Myr prior to
triggering; otherwise triggering wouldn't be needed, and a large
spread in stellar ages would result, which is not observed in
the youngest group, Upper Sco (Preibisch
\& Zinnecker 1999). The evidence of Table 1 suggests that it is
unlikely that such large masses of molecular gas would have remained
inert for such a long time.  The non-local flow picture avoids this
problem by making the triggering of cloud formation the same event as
the triggering of star formation.  

Putting it another way, one clear indication of triggering cloud and
star formation would be to see swept-up, moving gas producing stars.
The small dispersion in stellar proper motions in Sco OB2 provides
little evidence that swept-up gas driven by the oldest stars (Upper
Cen-Lupus) was decisive in producing star formation in the younger
Upper Sco-Oph regions.  These issues should be addressed further with
hydrodynamic simulations.

In summary, while driving from local star formation may help trigger
new star formation locally, large-scale structures with small velocity
dispersions and young stellar populations generally require
large-scale, external triggering. 

\section{Rapid formation of molecular clouds and stars}

The simulations shown by BHV suggest that clouds in the interstellar
medium may be formed ``rapidly'' by large-scale flows.  Specifically,
BHV showed that the flows could produce clouds which evolve to high
densities over scales of tens of pc nearly simultaneously (i.e.,
within a few Myr). These simulations did not follow molecular gas
formation or demonstrate gravitational collapse because they were
limited to densities $ < 100 \, \cm^{-2}$. As most of the gas at the
solar circle is in diffuse H I, it is necessary to consider the
transformation between atomic and molecular gas. A more complete
picture can be developed by augmenting the results of simulations with
some additional physical considerations, as described below.  In this
section we focus on conditions particularly relevant to low-density
star-forming regions in the solar neighborhood. 

\subsection{Cloud accumulation and molecular gas shielding}

Even with the typical ISM flow velocities $\sim 10$~$\kms$ found by BHV,
it can take tens of Myr to accumulate enough mass from the diffuse
interstellar medium to form a molecular cloud complex (\S 4).
However, a necessary (though not sufficient) condition for the
existence of molecular material in the solar neighborhood is that it
have a large enough column density to effectively shield H$_2$ and CO
from the dissociating ultraviolet radiation of the diffuse galactic
field.  This requires a minimum column density in hydrogen atoms of
roughly (van Dishoeck \& Black 1988; van Dishoeck \& Blake 1998) 
\be
N_H (min) ~\approx~ 1 - 2 \times 10^{21} \cm^{-3} \,, \label{eq:nhmin}
\en
or
\be
A_V (min) ~\approx~ 0.5 - 1\,.  \label{eq:avmin}
\en
Thus, even if the process of building up material from diffuse H I
takes a long time, the relevant lifetime for the {\em molecular} cloud
(or a dark cloud) at the solar circle
only begins once this minimum column density is attained. A
substantial portion - in some cases the majority - of the time spent
in adding mass to an {\em eventual} molecular cloud may not contribute
to the molecular cloud ``lifetime''.  Since it is much more difficult
to detect concentrations of atomic hydrogen in the galactic plane than
it is to find molecular clouds, any possible pre-molecular state of
the cloud would be essentially ``invisible''.  (See, for example,
Figure 2 of BHV, which illustrates the difficulty of finding atomic
gas associated with Taurus, even at its high galactic latitude.) 

Conversely, disrupting the molecular cloud may not always require
physically moving the gas large distances.  Instead, simply expanding
the cloud material to the point where the column density falls below
the critical value of $N_H (min)$ for self-shielding could be
sufficient, especially in low-density regions. We return to the
question of dispersal mechanisms in \S 5. 

Dame (1993) estimated that the ratio of H~I to H$_2$ 
within about 1 Kpc of the Sun is about 4:1.
If we assume that this ratio represents the
average relative timescales for molecular and atomic phases, and if
molecular regions last for 3-5 Myr as suggested by Table 1, the atomic
phase between cloud formation epochs might last $\sim 12-20$~Myr,
which is consistent with a substantial accumulation period as
atomic gas (at the solar circle).

\subsection{Molecular gas formation}

Rapid formation of molecular gas also requires a minimum density as
well as shielding. We will present detailed calculations of the
physical and chemical post-shock evolution of gas produced by
colliding flows in the diffuse neutral atomic medium in a subsequent
paper (Bergin \etal 2001).  For present purposes we can constrain
parameters as follows. For atomic hydrogen densities greater than
about $n_H \sim 10^{2} \cm^{-3}$, and pressures $(P/k)_4 \gtrsim 1$
(where $(P/k)_4$ is measured in units of $10^4 \cm^{-3} \K$), the
heating and cooling rates in an unshielded atomic medium are
sufficiently fast to approach temperature equilibrium in $\lesssim
10^5$~yr, which is essentially instantaneous for our purposes (see
Wolfire \etal (1995), their equation (10), and associated discussion).
We may then use the equilibrium results of Wolfire \etal (1995) to
estimate the temperature and density of the post-shock cooling layer
{\em prior} to the time at which shielding by dust (and self-shielding
by $\h2$) becomes important.  For pressures $(P/k)_4 \gtrsim 1$, the
temperature approaches values $\sim 30$~K for densities $\gtrsim 300
\, \cm^{-3}$. Because the temperature will decline further once
shielding becomes important, and cosmic ray heating will tend to
maintain a minimum gas temperature $T \gtrsim 10$~K, we expect
relevant gas temperatures to lie in the range 10-30 K. 

The dust temperature may also be an important factor in $\h2$
formation. Tielens and Allamandola (1987) show that the evaporation
time of the H atom from a grain becomes shorter than the timescale for
an H atom to scan a grain surface at temperatures $\gtrsim 30$~K.  The
exact temperature dependence is uncertain as it will depend on various
grain properties and on whether the grain itself is coated by a layer
of water molecules. Some detailed calculations by J. Black (2001,
personal communication) suggest that grain temperatures must be $T
\lesssim 15$~K in order for $\h2$ formation to proceed rapidly.  Dust
temperatures in the diffuse (unshielded) interstellar medium are
uncertain; observational estimates range from about 13 to 22 K
(Legache \etal 1998; Wright \etal 1991; Sodroski \etal 1997).

Some small amount of extinction therefore may be required to lower
grain temperatures such that H atoms remain on the surface long enough
to locate another H atom and react.  Burton, Hollenbach, \& Tielens
(1990) find that, for standard interstellar radiation fields, the dust
temperature should scale roughly as 
\be
T~\propto~ [ \, {\rm exp}\,  (-1.8  A_V /1.086)\,  ]^{0.2}\,.
\en
If the unshielded dust temperature is $\sim 20$~K, then shielding of
$A_V \sim 0.8$ would reduce the dust temperature to $\sim 15$~K. Thus
it seems reasonable therefore to assume that in the shielded
post-shock layer, dust temperatures will be low enough for efficient
$\h2$ formation. 

With this assumption, we can estimate the $\h2$ formation rate. If we
adopt the formula of Hollenbach, Werner, \& Salpeter (1971), 
\be
R_{HWS} ~=~ 2.25 \times 10^{-18} \, T^{1/2} \, y_f \, \cm^{-3} \, {\rm
s^{-1}}\,, \
\label{eq:HWS}
\en
where $T$ is the gas temperature, and assume a sticking fraction $y_f
= 0.3$, the formation timescale is 
\be
t_{HWS} ~\sim~ ( R_{HWS} \,  n_H )^{-1} ~\sim~ 15 \, n_3^{-1}
T_{10}^{-1/2} \, {\rm Myr} ~=~ 15 \, [ (P/k)_4]^{-1} T_{10}^{1/2}\,
{\rm Myr}\,, 
\en
where $n_3$ is the hydrogen density in units of $10^{3} \cm^{-3}$ and
$T_{30}$ is the temperature in units of 30 K. This timescale is long
compared to the timescales implied by stellar population ages in
molecular clouds, especially for pressures comparable to the typical
galactic pressure $(P/k)_4 \sim 1$ (Mathis 2000; see estimates
summarized by Norman 1995).  However, other sources yield differing
rates.  For example, the observational results of Jura (1975)
suggested net rates of formation approximately a factor of 3-10 higher
than implied by (\ref{eq:HWS}).  Similarly, the rates of Tielens \&
Hollenbach (1985), which were used by Koyama \& Inutsuka (2000) to
study $\h2$ formation in an unshielded post-shock gas, would yield
timescales about a factor of three shorter, 
\be
t_{HT} ~\sim~ 5 \,  n_3^{-1} T_{10}^{-1/2} \, {\rm Myr} ~=~ 5 \,
[(P/k)_4]^{-1} T_{10}^{1/2}\, {\rm Myr}\,. 
\en
(equivalent to setting $y_f \sim 1$ in (\ref{eq:HWS})).

It may be that cloud formation in the solar neighborhood is mainly
driven by pressure forces a few times larger than the average
pressure, so that $\h2$ formation is more rapid.  In addition, as we
show in the next section (\S 3.3), once a shielding column density is
attained, gravitational forces start to become important, which will
cause the cloud to contract and become denser; factors of a few
increase in density would be sufficient to ensure rapid molecular gas
formation.  Turbulent and clumpy internal structure (Elmegreen 2000)
may also play a role in elevating local densities and thus help form
molecules.

We will present a detailed analysis of molecular gas formation in a
subsequent paper (Bergin \etal 2001).  For the present, it appears to
be possible to form $\h2$, with CO formation following closely
thereafter (Bergin, Langer, \& Goldsmith 1995), within a few Myr once
gas densities reach $\sim 10^3 \cm^{-3}$ (see also Koyama \& Inutsuka
2000). 

\subsection{Gravitational instability} 

The observations imply that soon after molecular clouds are formed,
stars are produced. We next show that, under ``ideal'' circumstances,
clouds with sufficient shielding can collapse gravitationally, on a
sufficiently short timescale, and then consider limiting factors.

In our picture, the molecular cloud is the post-shock region of
converging flows. As described in \S 3.2, for gas pressures near the
fiducial value, the temperature in the post-shock region should
rapidly decay to values of $< 30$~K. An idealization of this
situation, which provides the most favorable conditions for
gravitational collapse, is an infinite, planar, isothermal, and nearly
static layer (since the post-shock flow velocity is greatly reduced at
high density). For such  a layer, the central pressure is (Ledoux 1951; 
Spitzer 1978; Elmegreen \& Elmegreen 1978)
\be
P_c ~=~ P_e ~+~ \pi G \Sigma^2/2 \,, \label{eq:pcgrav}
\en
where $P_e$ is the external pressure and $\Sigma$ is the total column
density through the sheet. Assuming that the cloud is molecular, the
internal pressure due to self-gravity is comparable to the external
pressure for a column density of hydrogen atoms 
\be
N_{H} ~\sim ~ 1.5 \times 10^{21} \, (P_e/k)_4^{1/2}\, \cm^{-2}\,,
\label{eq:nc} 
\en
or
\be 
A_V ~\sim ~ 0.8 (P_e/k)_4^{1/2}\,, \label{eq:avc}
\en
Thus, the column density needed to produce a detectable dark cloud,
and to allow molecular gas to form (\S 3.1, 3.2) is comparable to the
that required for self-gravity to be important in comparison with
external pressure forces (e.g., Elmegreen 1991). We argue that this
coincidence between the column density needed for molecular shielding
and that required for important self-gravitating forces is the basic
reason why star formation is presently occurring in virtually all
molecular cloud complexes of significant size in the solar
neighborhood. 

We next investigate the possible scales of gravitational instability
and the associated collapse timescales. For an isothermal infinite
sheet in hydrostatic equilibrium, neglecting the external pressure,
the critical spatial wavenumber for stability is (Ledoux 1951; Spitzer
1978) 
\be
k_c ~=~ \pi G \Sigma / c_s^2 \,,
\en
corresponding to a Jeans length
\be
\lambda_c ~\equiv~ {2 \pi /  k_c} ~=~ 0.7 \, T_{10} N_{21}^{-1}\, {\rm
pc}\,, 
\en
where we have assumed that the gas is molecular, $T_{10}$ is the
temperature in units of 10 K, and $N_{21}$ is the molecular hydrogen
column density in units of $10^{21} \, \cm^{-2}$. The maximum growth
rate $\Gamma_{max}$ occurs on a wavenumber approximately twice
critical, or a wavelength $2 \lambda_c$, and has a value (Simon 1965) 
\be
\Gamma_{max} ~\approx~ 0.67 \, \pi G \, \Sigma / c_s ~=~ 3.6 \times
10^{-14} \, N_{21} \, T_{10}^{-1/2}\,  {\rm s^{-1}}\,, 
\en
for a characteristic growth time
\be
\tau_{min} ~=~ \Gamma_{max}^{-1} ~\sim~  0.9 \times 10^6 \,
T_{10}^{1/2} \, N_{21}^{-1}\, {\rm yr}\,. 
\en
Linear growth rates remain within a factor of about two of the maximum
rate $\Gamma_{max}$ for wavelengths between about $\sim 1.07
\lambda_c$ and $\sim 15 \lambda_c$ (Simon 1965), so the above
timescale for collapse is not very sensitive to the scale
involved. Thus, once the column density approaches the minimum
shielding value, and the temperature drops below 30 K, it is possible
for molecular gas to collapse gravitationally on timescales of order
1-3 Myr over a wide range of length scales and masses. 

The criterion (\ref{eq:nc}) or (\ref{eq:avc}) is not a strict guide to
the onset of gravitational collapse.  In principle, gravitational
instability can occur at lower surface densities (e.g., Elmegreen \&
Elmegreen 1978).  However, low-surface density clouds are much more
susceptible to disruption by external pressures, and are more likely
to be supported against gravity by turbulent motions. To examine the
effects of external pressure distortion, consider an idealized
situation in which the flows produce a curved, expanding shell (i.e.,
Figure 3). An expanding shell can be stabilized against gravity if
(Vishniac 1983) 
\be
\Gamma_{max} ~\ll~ V_s / R_o\,,
\en
where $V_s$ is the shell expansion velocity and $R_o$ is the radius
(or characteristic radius of curvature) of the shell.  This implies
that the above sheet can be prevented from collapsing if the
characteristic radius of curvature is 
\be
R_o \ll 9 \, \left ( {V_s \over 10 \kms} \right ) \, T_{10}^{1/2}
N_{21}^{-1}\, {\rm pc}\,. 
\label{eq:exp}
\en
Because observed cloud structures tend to be much larger than a few pc
(and tend to be driven over much larger scales than this; \S 4),
expansion is unlikely to be a general obstacle to gravitational
collapse for column densities near or above the shielding constraint. 

Similarly, uniform rotation can also suppress gravitational
instability above a critical value of the Toomre $Q$ parameter 
\be
Q_c ~=~ {c_s \Omega \over \pi G \Sigma} ~=~ 0.338 \, \label{eq:glbq}
\en
(Goldreich \& Lynden-Bell 1965a), where $\Omega$ is the angular
frequency. \footnote{For differentially-rotating regions, shearing
perturbations can grow by large factors even when non-shearing
perturbations are stable (Goldreich \& Lynden-Bell 1965b; Toomre
1981).  However, in the case of shear, replacing $\Omega$ by the
epicyclic frequency $\kappa$ in (\ref{eq:glbq} and \ref{eq:oc})
changes the stability criterion only by factors of order unity.} In
terms of a maximum rotational velocity gradient, 
\be
\Omega_c^{-1} ~\sim~ 0.55 {\kms} \, T_{10}^{-1} \, N_{21}\, {\rm
pc^{-1}} \,. 
\label{eq:oc}
\en
Again, this does not appear to be a major limitation for column
densities near or above the necessary shielding length.  For example,
the radial velocity gradient across the main component of the Taurus
complex is $\approx 0.25$~$\kms \, {\rm pc^{-1}}$ (Kleiner \& Dickman
1985), although in some smaller regions the gradients may approach
$1$~$\kms \, {\rm pc^{-1}}$ (Arquilla \& Goldsmith 1986). 

\subsection{Turbulence}

It has been generally accepted since Chandrasekhar \&\ Fermi (1953)
that turbulent motions can give rise to internal pressure support,
which could prevent cloud and star formation. For example, if the
$\sim 2$~$\kms$ ``turbulent'' velocities observed in the Taurus
molecular complex corresponded only to extremely small-scale motions,
the resulting internal turbulent
pressure would prevent the post-shock gas from condensing to densities
$>  15 \, (P/k)_4\, \cm^{-3}$. This would imply that to form molecular
gas, either (1) pressures must be two orders of magnitude larger than
typical interstellar pressures, (2) turbulent motions damp rapidly, or
(3) these motions do {\em not} correspond purely, or even mostly, to
very small-scale turbulence. In fact,
Sasao (1973) demonstrated that large scale turbulence plays an
important role as a generator of astronomical objects, and that the
role of the turbulent pressure might be only of a higher
order. Moreover, based on numerical simulations, 
L\'eorat, Passot, \&\ Pouquet (1990), and Klessen et al. (2000) have
shown that local collapse may be hindered only if turbulence is
present at the very smallest scales; in a realistic turbulent medium,
turbulence can support the cloud globally while promoting local
collapse (Klessen et al. 2000).

We argue that, even in situations where (1) is not appropriate,
factors (2) and (3) can result in rapid formation. Recent numerical
simulations show that turbulent motions do decay rapidly in molecular
clouds, generally on a crossing time for any scale involved (Stone
\etal 1998; Mac Low et al. 1998;  Padoan \&\ Nordlund 1999), or on
timescales smaller than the free fall timescale (Mac Low 1999). Here
the important point is that the relevant distance for the crossing time
is the shortest dimension, not the longest.
For the fiducial pressure $(P/k)_4 = 1$, and a temperature of 10 K,
the physical length at a column density of $N_{21} = 2$ is $\sim
2/3$~pc, and the crossing time at a turbulent velocity of 1-2~$\kms$
would therefore be less than 1 Myr.

Star-forming molecular clouds do exhibit substantial turbulent
velocities of $\sim 1-2$~$\kms$ on size scales of a few pc (e.g.,
Larson 1981), so that not all of the turbulent motions have decayed at
the epoch of star formation. However, the internal motions of
molecular clouds most likely do {\em not} correspond to only very
small-scale turbulence, but instead contain substantial energy on
large scales, i.e., diverging and converging flows, as has been
suggested based on results of numerical simulations of cloud
turbulence by BVS.  These large-scale motions are particularly
susceptible to efficient dissipation of energy in shocks.  In
addition, simulations show that superposition of independent regions
along the line of sight cannot be ignored when attempting to interpret
observations (Kwan \& Sanders 1986; BVS). For this reason observed
turbulent velocities can easily result in an overestimate of internal
(small-scale) velocity dispersions, as has been pointed by Ostriker,
Stone, \& Gammie (2001); and it is difficult to determine when an
observed clump is actually a physical entity or the superposition
of different regions along the same line of sight
(Ballesteros-Paredes \&\ Mac Low 2001).  In other words, individual
clumps could have low internal velocity dispersions even though moving
supersonically relative to each other along the line of sight (Kwan \&
Sanders 1986). 

Moreover, Ballesteros-Paredes (1999, 2000) has shown that even if
there is equipartition between the kinetic, magnetic and gravitational
energy components (Ballesteros-Paredes \&\ V\'azquez-Semadeni 1995,
1997), the kinetic energy term (pressure) does not necessarily
contribute support against gravity; cloud compression as well as
expansion or disruption can result from such turbulence. In examining
these possibilities it is important to consider not only the internal
motions, but also surface forces as well (BVS). In fact, as has been
shown by Klessen \& Burkert (2001) and Klessen et al. (2001), large
scale turbulence is able to form dense, elongated structures that will
collapse rapidly, producing clustered star-forming regions. 

Our picture, then, of a molecular cloud is one in which the small
scale turbulent motions are smaller than frequently estimated because
of superposition effects, and which in any event are rapidly damped,
allowing the gas to reach high enough densities to form $\h2$ and CO
rapidly ($t < 10$~Myr). It is important to keep in mind that the low
efficiency of star formation in most nearby clouds (e.g., Cohen \&
Kuhi 1979) requires that only a small fraction (generally, a few
percent) of the molecular gas collapses to form stars.  Thus, damping
of turbulence need not be complete in the entire
cloud complex for our picture to hold.

\subsection{Magnetic fields: compression and the ``flux
problem''}\label{magnetic_effects} 

The pressure from (non-turbulent) magnetic fields potentially could
prevent post-shock densities from rising to high enough values to form
molecular gas rapidly. In ideal MHD, steady one-dimensional flow, the
shock relations for the field component $B_t$ perpendicular to the
shock front result in $B_{t} \propto \rho$ in the post-shock gas
(McKee \& Hollenbach 1980). Assuming an oblique shock, as illustrated
schematically in the left-hand panel of Figure 4, the field thus
becomes increasingly parallel to the shock front as the gas cools and
becomes denser, eventually limiting the maximum density achieved. The
ratio of the maximum post-shock density, $n_m$, to the initial density
$n_o$, is given by (McKee \& Hollenbach 1980) 
\be
{ n_m \over n_o} ~\sim ~  35 \, (P_e/k)_4^{1/2} \,
{B_{o,t} \over 1 \mu{\rm G}}\,, \label{eq:compress}
\en
where $B_{o,t}$ is the initial transverse magnetic field
strength. Assuming diffuse interstellar gas densities of a few
$\cm^{-3}$, flow velocities of order 10~$\kms$, and random magnetic
field strengths of $\sim 5 \, \mu$G (Mathis 2000), it would be
impossible to achieve the densities of $10^3 \cm^{-3}$ needed to form
molecular gas rapidly unless the flows are essentially parallel to the
magnetic field (so that $B_{o,t} < 1 \, \mu{\rm G}$) (e.g., Hennebelle
\& P\'erault 1999). Unless the magnetic fields are very strong, so
that flows are channeled completely along field lines (Passot,
Vazquez-Semadeni, \& Pouquet 1995 $=$ PVP; Ostriker \etal 1999, 2000),
cloud formation would seem to be a very unlikely event. 

However, the numerical simulations of PVP, Ostriker \etal (1999,
2001), and Heitsch \etal (2000) show that the geometry indicated in
Figure 4 and expressed in equation (\ref{eq:compress}) generally is
not relevant for understanding cloud formation when the magnetic field
is weak or of intermediate strength with respect to the turbulent gas
pressure.  Unlike the field geometry shown in the left hand panel of
Figure 4, clouds tend to form at bends or ``kinks'' in the magnetic
field (Figure 5; see also BVS). This means that there are regions in
the cloud where $B_{t} \rightarrow 0$ (approximately parallel to the
major axis of the cloud), and it is here where gas compression can
proceed unabated by magnetic forces. Thus the compression of the
parallel magnetic field in such configurations can delay, but cannot
ultimately prevent, post-shock gas from compressing to high densities
as it cools.

A more difficult question is whether the magnetic field component more
or less perpendicular to the shock front(s) and the main axis of the
cloud is strong enough to delay or suppress gravitational collapse
into stars.  A very large literature exists which assumes that
magnetic fields are initially strong enough to prevent gravitational
collapse (e.g., Mouschovias 1991, and references therein), although
recent observational results (Crutcher 1999; Jijina, Myers, \& Adams
1999; Lee \& Myers 1999; Bourke \etal 2001) and theoretical analyses
(Nakano 1998; Padoan \& Nordlund 1999; Ciolek \& Basu 2001) suggest
that the effects of magnetic fields in preventing or slowing collapse
may be much less than previously thought. 

To simplify as much as possible, consider our thin (infinite,
isothermal, self-gravitating) sheet from \S 3.1-3.3, threaded by a
perpendicular magnetic field.  Gravitational collapse can ensue only
if (Nakano \& Nakamura 1978) 
\be
G \Sigma_c^2  ~>~ { B^2 \over 4 \pi^2 } \,. \label{eq:sigmacrit}
\en
Multiplying both sides by the area of the cloud, and taking the square
root,
\be
(4 \pi^2 G)^{1/2} M_c ~>~ \Phi_B \,,  \label{eq:phi}
\en
where $M_c$ is the cloud mass and $\Phi_B$ is the magnetic flux
threading the cloud. Clouds satisfying the relation
(\ref{eq:sigmacrit}) or (\ref{eq:phi}) are said to be magnetically
supercritical; otherwise, the clouds are magnetically subcritical. (In
different geometries the numerical relation in (\ref{eq:phi}) changes
modestly, without altering the fundamental relation.) If flux-freezing
holds, a subcritical cloud will never be able to collapse
gravitationally.  Magnetic flux therefore must be removed from the
cloud before stars can form.  This cannot occur easily in the diffuse
interstellar medium or even in low-column-density regions of molecular
clouds (Myers \& Khersonsky 1995), because the magnetic field lines
are well-coupled to the gas.  In this situation, collapse to stars
cannot occur until magnetic flux is removed via ambipolar diffusion in
the dense, highly-shielded cloud regions where the ionization is very
low.

In the standard model of low-mass star formation, protostellar clouds
can be magnetically subcritical by a wide margin (e.g., Shu, Adams, \&
Lizano 1987; Mouschovias 1991, and references therein).  The timescale
for the necessary ambipolar diffusion of magnetic flux can be as long
as 5-10 Myr, depending upon how subcritical the cloud is
initially. But this is inconsistent with the stellar population ages.
One would expect a delay of at least several Myr between the time of
molecular gas formation and the onset of star formation (Palla \&
Galli 1997), and this is not observed in the solar neighborhood.  In
addition, it is implausible that all regions of all clouds are
subcritical to the same extent; therefore there should be {\em spread}
of diffusion timescales, and thus ages, comparable to the overall
diffusion time.  This would result in age spreads of 5-10 Myr in the
case of very subcritical initial conditions, which again is not
observed.

The solution to this dilemma is that molecular clouds must be
initially supercritical (Hartmann 1998; Nakano 1998), or at least
close to critical so that only a small amount of magnetic flux need be
diffused away (e.g., Ciolek \& Basu 2001). Reassuringly, observations
indicate that molecular clouds are nearly critical or slightly
supercritical on large scales (McKee 1989; McKee \etal 1993), and even
in denser cloud cores (Crutcher 1999; Bourke \etal 2001).

A clue to why clouds are initially supercritical is given by numerical
simulations.  Ostriker \etal (2001) noted that condensations can form
whether the regions inside their computational region are magnetically
sub- or supercritical, but gravitational collapse ensues only for the
supercritical cases. PVP also found that gravitational collapse can
occur when the computational region (in this case a large section of
the galactic disk in 2 dimensions) is magnetically supercritical.
These results demonstrate the importance of boundary conditions. With
periodic boundary conditions (as is the case of the numerical
simulations performed to date by PVP, Ostriker \etal 2001; Padoan
\etal 2001; and Heitsch, Mac Low \&\ Klessen 2001) the total mass is
fixed. Then, starting with roughly uniform magnetic fields and
densities, an initially subcritical box will always be subcritical
(and the clouds it forms will be subcritical) in absence of diffusion
or reconnection. As emphasized by Heitsch et al. (2001), protostellar
collapse is inhibited by magnetic fields if they initially provide
magnetohydrostatic support; otherwise, they will slow, but will not
stop, collapse.

For a given average magnetic field strength and gas density, the size
of the computational region then determines whether the region is
subcritical or supercritical. The key parameter is the ``accumulation
length'' (e.g., Mestel 1985), the distance along a magnetic flux tube
needed to achieve the critical column density. The accumulation length
$l$ for forming a magnetically critical cloud is roughly 
\be
l_c ~\sim~  430 \, (B/5 \mu G) \, (n_H/1 \cm^{-3})\, {\rm pc}\,,
\label{eq:lc} 
\en
where the fiducial values are typical for the galactic interstellar
magnetic field and hydrogen density in the vicinity of the Sun (Heiles
1995; Mathis 2000; Beck 2001).  Thus, for typical ISM values of gas
density and magnetic field strength, computational regions larger than
about 400 pc will be supercritical as a whole in the absence of
diffusion or reconnection. 

It seems at least intuitively plausible that the computational ``box''
for cloud formation should be at least as large as some relevant
dimension perpendicular to the galactic plane, if not larger.  A
region of at least $\sim 270$~pc would be required to match one scale
height above and below the plane in the atomic hydrogen distribution
(Mathis 2000; Dickey \& Lockman 1990).  A scale of this length is also
strongly suggested by the Orion molecular complex, which extends about
140 pc below the galactic plane.  Another important constraint is the
overall pressure scale length.   Boulares \& Cox (1990) have argued
that the magnetic pressure in the local interstellar medium at the
solar radius drops by only a factor of two at distances of $\pm
400$~pc above and below the plane. Thus, a computational volume with a
lateral dimension of one vertical pressure scale height would be
approximately magnetically critical at the typical densities and field
strengths in the solar neighborhood. 

These considerations suggest that the accumulation lengths for
molecular clouds are simply large enough that clouds are supercritical
when formed. A large accumulation length is also consistent with the
results of numerical simulations (PVP; BHV; \S 4), in which clouds are
formed from flows extending over several hundred pc.  It may not be a
coincidence that the formation of a $10^6 \msun$ giant molecular cloud
out of diffuse material at $1\, \cm^{-3}$ requires the accumulation of
material from a volume of $\sim 400$~pc in size (Williams \etal 2000).

The possibility that clouds {\em can} have large enough accumulation
lengths to be magnetically supercritical does not mean that they {\em
will} be supercritical. However, we argue that subcritical clouds
generally are not molecular. 
The internal 
magnetic pressure -- random plus ordered components -- must be less
than or at most comparable to the external turbulent pressure in the
interstellar medium; otherwise, the cloud would expand. Assuming again
a sheet-like geometry, the {\em tangential} external pressure (along
the sheet) needed to confine a magnetic field $B$ oriented roughly
perpendicular to the sheet is  
\be
P_t~ \gtrsim {B^2 \over 8 \pi} \,, \label{eq:bpe}
\en
Now $P_t$ will be at most equal to, and more likely less than, the
pressure component $P_e$ {\em normal} to the sheet (otherwise the
sheet would become compressed in the opposite direction).  But column
densities satisfying equation ({\ref{eq:nc}) imply $G \Sigma^2 / 2
\gtrsim P_e$.  Thus, when ({\ref{eq:nc}) is satisfied, we have 
\be
G \Sigma^2 / 2 ~ \gtrsim P_e ~ \gtrsim P_t ~\gtrsim~  B^2  /(8 \pi)\,,
\label{eq:gpb} 
\en
which satisfies (\ref{eq:sigmacrit}).  Thus the cloud tends to be
supercritical at column densities high enough for molecular gas
formation. 

The reason why previous considerations of (strongly)
magnetically-subcritical clouds do not come to this conclusion is that
they do not apply a criterion such as equation (\ref{eq:bpe}), and so
the magnetic field can be of arbitrary strength. Because the internal
magnetic pressure of a subcritical cloud exceeds the force of gravity,
such a cloud must expand unless it is confined by external pressure
forces (Fiedler \& Mouschovias 1993); and it is generally unlikely
that external pressure forces can hold a strongly-subcritical cloud
together (Nakano 1998; Hartmann 1998), particularly if these pressures
are turbulent and therefore highly anisotropic and time-dependent
(BVS). 

It should be emphasized that the above discussion deals mainly with the
large scale.  In principle it is always possible to find a small enough
scale within a supercritical cloud that is subcritical.  However, given
the importance of boundary conditions as described in the preceding
paragraph, we think that the large scale is controlling.  If supercritical
collapse can proceed on the scale of the cloud, subsequent fragmentation 
in principle can occur to produce smaller supercritical cores.

\subsection{Simulation of cloud formation}

To support the picture of cloud formation outlined above, with
particular emphasis on the issue of magnetic field support, we
consider the behavior of the galactic interstellar medium from PVP to
help support these assumptions. To this end, we display results from a
run called r28cfa, which has the same parameters than the fiducial run
28 in PVP, but performed on a workstation at CfA\footnote{In BHV (also
Fig.~\ref{vbrho}) we showed a simulation, called Run 28.800, with
similar parameters but with higher spatial resolution ($800^2$
pixels), in which star-formation was turned off after 65~Myr.}. 

The simulations consider a two-dimensional section of the interstellar
medium in the galactic plane at the solar circle, which has dimensions
of 1000 by 1000 pc, large enough to encompass a critical accumulation
length (\S 3.4.2). The model includes self-gravity, magnetic fields,
coriolis force, galactic shear, diffuse heating, cooling, and stellar
energy input, as well as a scheme for star formation. Details of the
model are given in PVP. We note that (see Table~1 in PVP for the
standard parameters), the initial conditions are random in all
variables with phases uncorrelated. In particular, the magnetic field,
which is entirely in the plane of the simulation, has a uniform
initial component of $1.6~\mu$G, and a random component of $5~\mu$G,
corresponding to an initial r.m.s. magnetic pressure of $P_B/k \sim 7
\times 10^3$~$\cm^{-3} \K$. 
 
Small random perturbations in velocity and density are introduced
initially and then the system is left to evolve. Once the gas reaches
densities $\ge$ 30~cm\alamenos 3, and if the local velocity field is
convergent ($\nabla\cdot {\bf u} < 0$), star formation is assumed to
occur and an energy source corresponding to the energy of massive O
stars is \ turned on for a lapse of time of 6~Myr, the lifetime
typical for massive O stars (see VPP). 

This scheme for star formation has the following ``collateral''
effects. First of all, it prevents densities from getting much larger
than the threshold density for star formation. On the other hand,
while the energy input is point-like (i.e., at small scales), its
long-term effect (after several Myrs) the turbulence appears at all
scales. This is because the ``HII regions'' of hot gas formed by the
new-born stars expand, interact with each other, and form structures
at all scales.

Much, if not most, of the stellar energy input into the interstellar
medium arises from supernovae (Spitzer 1978; Wada \& Norman 2001).
The simulation presented here does not include such energy input
(although there exists a variation of this code that includes
supernovae, see Gazol-Pati\~no \&\ Passot 1999). In reality, much of
the driving flows are initially in hot, ionized bubbles, rather than
the relatively low-temperature diffuse atomic flows considered here
(e.g., McCray \& Kafatos 1987).  However, this should not change our
general conclusions, since the hot gas must eventually cool and become
atomic before it can make a transition to the molecular phase. 

The simulation starts at constant density (1~cm\alamenos 3) with
fluctuations of about 20\%. Because the perturbations are
uncorrelated, strong shocks appear and form the first low-density
clouds. In Figure \ref{acumula}, we show four snapshots of run 28 at
the $t=$0, 9, 64 and 117 Myr.  The grayscale indicates regions
of increased density; ``clouds'', defined as regions where the
density exceeds 15~cm\alamenos 3, are outlined by black contours.
The magnetic field directions are indicated by the arrows.

The first point to notice in Figure \ref{acumula} is that the first
dense structures formed (second frame) are separated by about 100 pc,
and so this corresponds to their accumulation lengths. Evolution from
densities of 1~cm\alamenos 3 up to densities of 30~cm\alamenos 3 takes
about 10 Myr; after that, relatively dense clouds are present most of
the time. To form the larger and denser clouds in the third and fourth
frames, accumulation has proceeded over larger distance scales. Figure
\ref{acumula} shows that after star formation has proceeded for
several tens of Myr, most lines of sight pass through few regions of
higher than average density.  The dominant flow is in the
$x$-direction (direction of galactic rotation); along any cut at
constant $y$, there no clouds, or only one.  Even considering
regions with lower densities than our ``clouds'', there are a
relatively small number of structures intercepting a given line at
constant $y$. This indicates that the scale of accumulation along the
$x$ direction is a large fraction of the 1 kpc box length, and
therefore it is not surprising that the clouds are supercritical.

The simulation shows that star formation is spatially and temporally
correlated; once star formation ensues in a given region, some of the
dispersed material recondenses nearby (several tens of pc distant) a
few to tens of Myr later (see also Elmegreen \& Efremov
1996). Although the simulation does not address the question directly,
we suggest that the disappearance of old clouds and the formation of
new clouds results from a combination of actually moving gas from one
place to another as well as dissociation and reformation of the
molecular material (see following subsection). This behavior will be
analyzed more carefully in a subsequent paper (Ballesteros-Paredes
\etal 2001).  Here we merely note that the simulations demonstrate the
important distinction between the star-forming history of a region
tens to hundreds of pc in size and the evolution of an individual
parcel of gas.  Over a sufficiently large volume, star formation can
proceed for tens of Myr even though the individual molecular cloud
regions form, produce stars, and disperse more rapidly (cf. Elmegreen
1979). 

The simulations support the point made in \S 3.4 that the deceleration
of large scale flows can result in a line-of-sight ``turbulent
velocity broadening'' which does not correspond to internal pressure
support, but rather to the {\em compression} of the cloud. Such
features may be seen in all the clouds in the simulation, as can be
seen also in BVS (see also the expanded view of a cloud in Figure
~\ref{vbrho}).

Figure \ref{fig:caja} shows the evolution of energies within the
entire computational region. The gravitational energy (solid line) has
the opposite sign to facilitate comparison with the other
energies. The magnetic and (internal) energies remain nearly constant
for the first 50 Myr of the simulation, decaying slightly thereafter.
The overall kinetic energy decreases slightly at first due to
dissipation, and then eventually steadies or even rises due to stellar
energy input.  The main evolution is in the gravitational potential
energy in the volume, which rises rapidly until it becomes roughly
comparable to the other energy terms.

Figure \ref{fig:nube} compares the evolution of the energies of the
``clouds'', defined as connected grid cells with densities greater
than $15 \, \cm^{-3}$. Each line in Figure~\ref{fig:nube} represents
the sum over all clouds of various energy components (gravitational,
thermal, kinetic or magnetic). (The kinetic energy for each cloud is
calculated in the frame of reference of the cloud itself, i.e., bulk
motion is not included.) Figure~\ref{fig:nube} shows considerable
evolution with time of all energy components.  The magnetic energy
tracks the kinetic and internal energies most closely, as predicted by
the pressure argument (\ref{eq:bpe}); this is made explicit in Figure
\ref{fig:pvst}, which converts the values in Figure \ref{fig:nube} to
cgs pressures by dividing the total energies of the clouds by their
total area (since the simulations are two-dimensional). Comparison
between Figures \ref{fig:nube} and \ref{fig:pvst} shows that much of
the energy evolution in Figure \ref{fig:nube} is due to increasing
cloud volume (area) rather than changing energies per unit volume
(area). The gravitational energies/pressures rise as time progresses, 
and they exceed the corresponding magnetic values as would be required 
by the argument leading to equation (\ref{eq:gpb}) for self-gravitating
clouds.

Figure \ref{fig:emvseg} compares the magnetic ($E_m$) and (negative)
gravitational ($E_g$) energies of individual clouds at three selected
times (the same last three timesteps in Fig~\ref{acumula}). Again, the
magnetic and gravitational energies are correlated, with the larger
clouds being dominated by gravitational energy and smaller clouds more
likely to be magnetically dominated.  This corresponds to the larger
clouds being supercritical, as expected from
\S\ref{magnetic_effects}.

In assessing criticality, we compare energies rather than use the
mass-to-flux relation (\ref{eq:sigmacrit}). However, virial theorem
arguments for flattened clouds give essentially the same result as the
force perturbation analysis for the sheet (Nakano \& Nakamura 1978;
Strittmatter 1966). Note that the mass-to-flux criterion relates $B$
to cloud mass $M$, whereas the energies in Figure (\ref{acumula})
involve squares of these quantities; thus, the results shown in Figure
(\ref{fig:emvseg}) correspond to clouds which are within factors of a
few of critical, as indicated by observations (e.g., Crutcher 1999).

It is possible that hyperviscosity used in the scheme helps remove
some magnetic flux and thus tends to make clouds in the simulation
more supercritical than they should be. However, the large
accumulation lengths of the clouds in the simulation compared to the
small scales on which the hyperviscosity matters (Vazquez-Semadeni,
Passot, \& Pouquet 1995) suggests that the latter is not the essential
factor in making supercritical clouds. This is corroborated by the
fact that the total magnetic energy over the entire simulation has a
very small decrease over the 130 Myr. In summary, the results of the
numerical simulations tend to support the idea that molecular clouds
will be supercritical due to large accumulation lengths.

\subsection{Dispersal}

It is evident that rapid cloud dispersal after the onset of star
formation is a crucial part of the explanation of why star forming
populations in molecular clouds are young. If massive (O) stars are
present, or if a nearby SN explosion has occurred, removing the
remaining gas on a few Myr is no problem (cf. Cep OB2, Patel \etal
1998; also the $\lambda$~Ori cluster, Dolan \& Mathieu 1999).
Dispersal of molecular clouds by low-mass stars is another matter.
The power of a group of low-mass stars to eject material is uncertain,
but is probably much smaller than that of a single O star.  Moreover,
the efficiency of star formation appears to be much higher in
high-density regions producing high-mass stars, and thus the low-mass
stars need to disperse much more gas relative to the total mass in
stars.

The outflows of young low-mass, pre-main sequence stars can eject
material, as long the ejection is not simply confined to a narrow,
highly-collimated jet, but exerts significant ram pressure over at
least a moderate solid angle (cf. Matzner \& McKee 1999, 2000).
Matzner \& McKee (2000) suggest that outflows from low-mass stars
might result in star-formation efficiencies of $\sim$~30\%; in other
words, these stars might be able to prevent twice their mass in
external gas from forming other stars. However, this level of mass
ejection would not explain the low efficiencies of a few per cent in
many star-forming regions (Cohen \& Kuhi 1979).

In general, the mass loss rates of T Tauri stars and FU Ori objects is
consistent with about 10\% of the accreted mass being ejected (Calvet
1991).  Let us assume that this ratio of mass ejected to accreted
characterizes the entire star formation process.  The outflows almost
certainly dissipate most of their energy in radiative shocks, so that
momentum conservation is the relevant consideration for dispersal of
molecular gas.  If the typical wind velocity is assumed to be $\sim
200$~$\kms$, then in principle an amount of mass $\sim 100$ times as
large as that ejected can be swept up in shells to velocities $\sim
2$~$\kms$, comparable to the escape velocities from typical molecular
cloud complexes.  This would imply a possible ejection of ten times as
much mass as formed into stars, or a star formation efficiency of
$\sim 10$\%.  This is slightly smaller than the Matzner \& McKee
(2000) estimate, but still higher than the $\sim$ 1-2 \% found in
Taurus and other regions.

In addition, unless the flows eject the molecular gas at $\approx
10$~$\kms$, gas cannot be dispersed spatially very far on timescales
of a few Myr. Essentially, this is the inverse of the formation
problem.  If ejection did occur at high velocities, the amount of
cloud mass that could be dispersed would be strongly reduced because
most of the kinetic energy of the flow would be dissipated in the
(radiative) shock as it sweeps up molecular material.

The reduction of shielding almost certainly plays an important role in
the ``disappearance'' of molecular gas near young, low-mass stars.
Molecular clouds are not spherical; they must be clumpy (e.g., Blitz
\& Shu 1980; Kwan \& Sanders 1986; Hollenbach \& Tielens 1999), and
often appear filamentary (e.g., Figure 1), or even fractal (see
Elmegreen 1997 and references therein). If the stellar energy input
can  expand the surface area of the dense gas sufficiently, the
reduction in shielding can dissociate the molecular gas. Actually,
Elmegreen (1997) estimated that radiation from an ionizing O star can
travel twice as far as otherwise would be expected if the clouds are
fractal, with a small filling factor. In the inverse process to cloud
formation, dispersal can result in turning the gas into atomic form so
that it is difficult to observe.  For regions like Taurus, where the
typical extinctions are $A_V \sim 1-2$ (e.g., Arce \& Goodman 1999),
an expansion of surface area of only a factor of two to three, with a
consequent reduction of column density by the same factor could
suffice to turn much of the molecular gas atomic (equation
{\ref{eq:avmin}).  In this case it is necessary to expand the gas by
only a few pc in a few Myr to make the molecular cloud
``disappear''. This effect is also important in high-mass regions, as
the photodissociating and photoionizing fluxes of massive young stars
will generally eliminate molecular gas long before the actual material
is dispersed to large distances.

Thus rapid dispersal of low-mass star-forming molecular clouds by
stellar winds is possible, though uncertain in its actual
efficiency. It may be that low-surface density molecular clouds are so
lightly bound gravitationally that the injection of even modest
amounts of stellar energy suffice to disrupt them.  Perhaps external
ram pressure forces also assist in injecting additional turbulent
energy which makes star formation efficiencies low to begin with.
Is it interesting that in Taurus, the molecular gas seems to extend
systematically west of the T Tauri stars (Figure 1). Is the gas being
blown systematically away from the stars in this direction?  In any
event, it is clear from several of the small groups in Taurus (for
instance, the groups at $b \sim -10$, $l \sim 178$, and the L1551
group at $b \sim -20$, $l \sim 179$) that the molecular gas is being
swept away rapidly due to some mechanism.  Perhaps large-scale
turbulent flows, in addition to forming clouds, may help disrupt the
least-tightly bound complexes. In any event, the problem of rapid
dispersal of molecular gas in low-mass, low-density remains one of the
biggest challenges to the picture presented here. 

\subsection{Summary outline of rapid cloud and star formation}

To explain the small age spreads (less than lateral crossing times) of
spatially-extended star-forming regions and stellar associations,
large-scale (hundreds of pc) flows must be involved in forming
molecular clouds (e.g., McCray \& Kafatos 1987; PVP; BHV).  These
flows may take tens of Myr to accumulate enough mass to form stars,
but the resulting clouds become molecular only when column densities
and volume densities reach threshhold values.  At these levels of
column density and volume density, gravitational forces are important
relative to external and internal pressure support. As shown in recent
numerical simulations, rapid dissipation of turbulence is likely to
allow gravitational collapse to occur, at least in restricted regions,
with timescales of a few Myr or less at threshhold column densities.
The large-scale nature of the flows lead to large accumulation
lengths, increasing mass-to-flux ratios; this feature, plus pressure
balance constraints, imply that clouds are generally both
self-gravitating and magnetically supercritical when they become
molecular, and thus ambipolar diffusion reduction of magnetic flux is
not an essential feature of star formation.

\section{Discussion} 

\subsection{Large-scale structures}

The probable importance of high-velocity flows driving cloud formation
in the interstellar medium is supported by the existence of molecular
gas structures which lie considerable distances out of the galactic
plane. The recent large-scale CO map of the galaxy by Dame, Hartmann,
\& Thaddeus (2001) gives the impression that the Orion A and B clouds
are part of a bubble extending below the galactic plane.  Whether or
not this is the case, some mechanism is required to move the gas as
much as 140 pc out of the galactic plane; driving by
stellar/supernovae energy input can naturally produce such structures
(see, e.g., Shapiro \& Field 1976, or more recent numerical
simulations by Avillez 2001 and Avillez \&\ Mac Low 2001). It has been
suggested that the Orion complex is the result of a high-velocity
cloud crashing into the galactic plane (e.g., Franco \etal 1988), but
flows from stars/SN are at least an equally probable, if not more
plausible, explanation. In this connection one wonders if the stars in
Gould's Belt (Poppel 1997) were not produced in clouds created by
complex interacting flow patterns moving material around and out of
the galactic plane, accidentally producing a structure which appears
coherent and tilted with respect to the plane. 

In this connection we note that Olano \& P\"oppel (1987) suggested
that the Taurus clouds might have formed as part of the general
expansion of the gas responsible for Gould's Belt.  However, the age
of the stellar population in Taurus (\S 2.1) is much smaller than the
age (18$\pm$3 Myr) suggested by Olano \& P\"oppel.  More generally,
our simulations of the interstellar medium (and others) demonstrate
the importance of large-scale flows (cf. Figure 5), so that the Olano
\& P\"oppel model of a shell snowplowing into a static medium is
highly unrealistic.  Due to the overlapping of long-range flows,
efforts to attribute cloud origins to single events or structures will
generally fail. 

The picture presented here, although envisioning cloud formation over
hundreds of pc, is basically a local one compared with spiral arm
structure.  Spiral arms clearly play a role in star formation by
accumulating gas, and in principle a spiral wave shock is just as good
as a supernova or stellar-wind-driven shock for compressing gas.  In
our picture, the energy put into the interstellar medium by massive
stars rapidly produces more complex structure within the spiral arm,
as for example in the picture outlined by Elmegreen (1979),
in which individual molecular cloud regions form, produce stars, and
disperse more rapidly within the general arm region.

An important part of our explanation of the rapidity with which stars
form after molecular gas formation is the notion that the flows are
initially atomic and only become molecular when sufficient column
density has been accumulated.  This picture of atomic to molecular
gas conversion is most appropriate near the
solar circle, where most of the gas is atomic; it may not be relevant
for inner regions in our Galaxy, where the molecular component
dominates, or more generally in other galaxies with different ratios
of atomic and molecular gas (Pringle, Allen, \& Lubow 2001).  However,
rapid star formation is likely to be appropriate in any case (Elmegreen 2000;
Pringle \etal 2001), although difficult to prove in the absence
of sufficiently precise stellar population ages.

\subsection{Supercritical star formation}

The picture we have outlined above is at variance with the standard
model of low-mass star formation.  The standard picture assumes that
stars form in dark clouds because it is only in such regions that the
ionization decreases to a level where ambipolar diffusion can proceed
(Shu \etal 1987; Bertoldi \& McKee 1996).  Instead, we suggest that
dark molecular clouds are the sites of star formation because they
represent a stage of cloud evolution closer to stellar densities than
represented by the atomic phase, partly as a result of turbulent
energy dissipation. Furthermore, we argue that molecular clouds form
in a magnetically supercritical state (\S 3.4.2) and therefore
ionization effects on ambipolar diffusion rates are generally not
important. 
 
Sometimes it is argued that, although molecular cloud complexes are
supercritical as a whole, star formation proceeds from subcritical
units within such cloud complexes (e.g., McKee \etal 1993).  It seems
more plausible to assume that gravitational collapse occurs first in
supercritical regions; subcritical regions may be disrupted by stellar
energy input before enough ambipolar diffusion can occur to permit
gravitational collapse (Hartmann 1998). 

\subsection{Efficiency and galactic star formation rates}

It follows from the above that the low rate of galactic star formation
is not the result of slowing by ambipolar diffusion of magnetic fields
through dense gas; instead, it is the result of a low efficiency in
converting gas into stars (Hartmann 1998; Elmegreen 2000).  The rapid
dispersal of clouds may be the main factor determining the efficiency.
It is much easier to see how turbulent flows powered by stellar energy
input would form, shape and disrupt clouds than trying to maintain a
quasi-equilibrium configuration which would allow clouds to survive
for long periods; and this is why, as shown by observations (\S 2),
clouds do not have long lifetimes. 

The idea that stellar ionization energy and winds are responsible for
limiting the efficiency of star formation initially seems to be in
conflict with the high star-forming efficiencies of dense regions
(Lada \& Lada 1991), which form disruptive high-mass stars. However,
stellar winds and supernovae can have a powerful impact on very large
scales, and it can be much easier to eject distant but low-density
gas, while very dense natal material can be much more difficult to
disperse, even if it is nearby. 
Cep OB2 (Patel \etal 1998) and other regions show how the large-scale
effects of stellar energy input can clear out atomic and molecular gas
over many tens of pc on timescales of 10 Myr which otherwise might
eventually collapse gravitationally if left undisturbed.  

\subsection{Protostellar collapse}

Another implication of our picture is that slow models of star
formation, where cores in hydrostatic equilibrium evolve slowly over
many dynamical timescales before collapsing, generally are not
appropriate. The flow-driven model predicts that, since the
newly-formed molecular cloud is almost immediately susceptible to
gravitational collapse, as well as being compressed by external flows,
it would be natural to observe systematic motions of contraction on
large scales.  The large-scale infall motions discussed by Williams \&
Myers (2000) could be a natural consequence of our model.   

Although protostellar cloud cores are often modelled as hydrostatic
structures, it is difficult to imagine conceptually how flow-driven
cloud formation would produce such static, slowly-evolving
structures. In fact, as BVS show, hydrostatic equilibrium can not be
obtained in turbulent flows until protostellar densities are
reached. Stars can be formed by dynamic processes and still rapidly
achieve hydrostatic equilibrium because their cooling times are many
orders of magnitude longer than their characteristic free-fall times;
thus the energy dissipated by shocks can be internalized and diffused
within a static interior.  Molecular clouds and cores do not share this
property. Shocks in the supersonic flows can and do dissipate
turbulence to relatively low levels, but true hydrostatic equilibrium
is highly unlikely.  Indeed, core statistics do not seem to suggest
that they are very long-lived entities (Lee \& Myers 1999). 

Protostellar cores generally are not spherical, and may even have a
tendency to be prolate\footnote{If not fractal, since the sizes of
these cores are only few times the beam size used in the
observations.} (Jijina, Myers, \& Adams 1999), which is much easier to
explain if cores are not in hydrostatic equilibrium (Fleck
1992). Jones, Basu, \& Dubinski (2001) recently have argued that cores
are generally triaxial, are closer to prolate, and thus closer to a
sustainable hydrostatic equilibrium than previously thought. However,
a dynamic picture would also naturally lead to an approximate
``triaxiality'' (i.e., all dimensions differing) without requiring any
special conditions. 
 
Therefore our flow-driven, rapid star formation picture is consistent
with the view that cloud cores are formed, at least in part, by the
collision of large-scale supersonic flows within molecular clouds
(Elmegreen 1993; PVP; BVS; Padoan \etal 2001), giving rise to
clustered star formation (Klessen \etal 2000; see also Elmegreen \etal
2000). 
 
\section{Conclusions}

The evidence of stellar populations shows that molecular clouds in the
solar neighborhood generally form rapidly, produce stars rapidly, and
disperse quickly, all within a timescale of only a few Myr.  In some
cases, the age spread of the young stars is much smaller than the
lateral crossing time.  
We have shown that this surprising behavior can be understood in the
context of cloud formation driven by large-scale flows in the
interstellar medium caused by global stellar winds and supernovae.
To understand the observations it is necessary to account for the
conversion from atomic to molecular gas and back again. The
observational requirement for rapid star formation can be satisfied
because the column densities necessary for formation of molecular gas
are comparable to those required for self-gravity to become important
in the solar neighborhood, and because collapse times can be as short
as 1 Myr under these conditions. Our picture requires that magnetic
fields, while having important dynamical effects, do not substantially
slow or prevent collapse in at least some portions of molecular
clouds; we have presented both general theoretical arguments and
numerical simulations in favor of this conclusion. 

While we have sketched a plausible explanation of the observations,
much more work remains before the picture of rapid star formation can
be placed on a firmer theoretical basis.  One issue is whether the
outflows from low-mass stars can effectively disperse low-density
star-forming regions on the required short timescales.  The rapid
dissipation of small-scale turbulence is required by our picture, and
while there is some current justification for this assumption, many
details remain uncertain.  Finally, the application of flow-driven
star formation to higher pressures and densities than considered here,
with ultimate application to understanding the formation of high-mass
stars and clusters, has yet to be undertaken. 

We are grateful to John Black for an extremely detailed and thorough
``very rough draft'' on the topic of $\h2$ formation.  We also thank
Mordecai Mac Low for comments on the manuscript and Ellen Zweibel for
useful conversations. Tom Megeath and Tom Dame graciously gave
permission to display Taurus $^{12}$CO results, and the latter
provided references on gas column densities in the solar
neighborhood. We also acknowledge a useful and thorough report from
the referee, Bruce Elmegreen. LH acknowledges the support from NASA
Origins of Solar Systems grant NAG5-9670. JB acknowledges support from
NASA Astrophysical Theory Program grant no. NAG5-10103, and from
CONACYT grant no. 88046-EUA.

\begin{figure}[ht]
\plotone{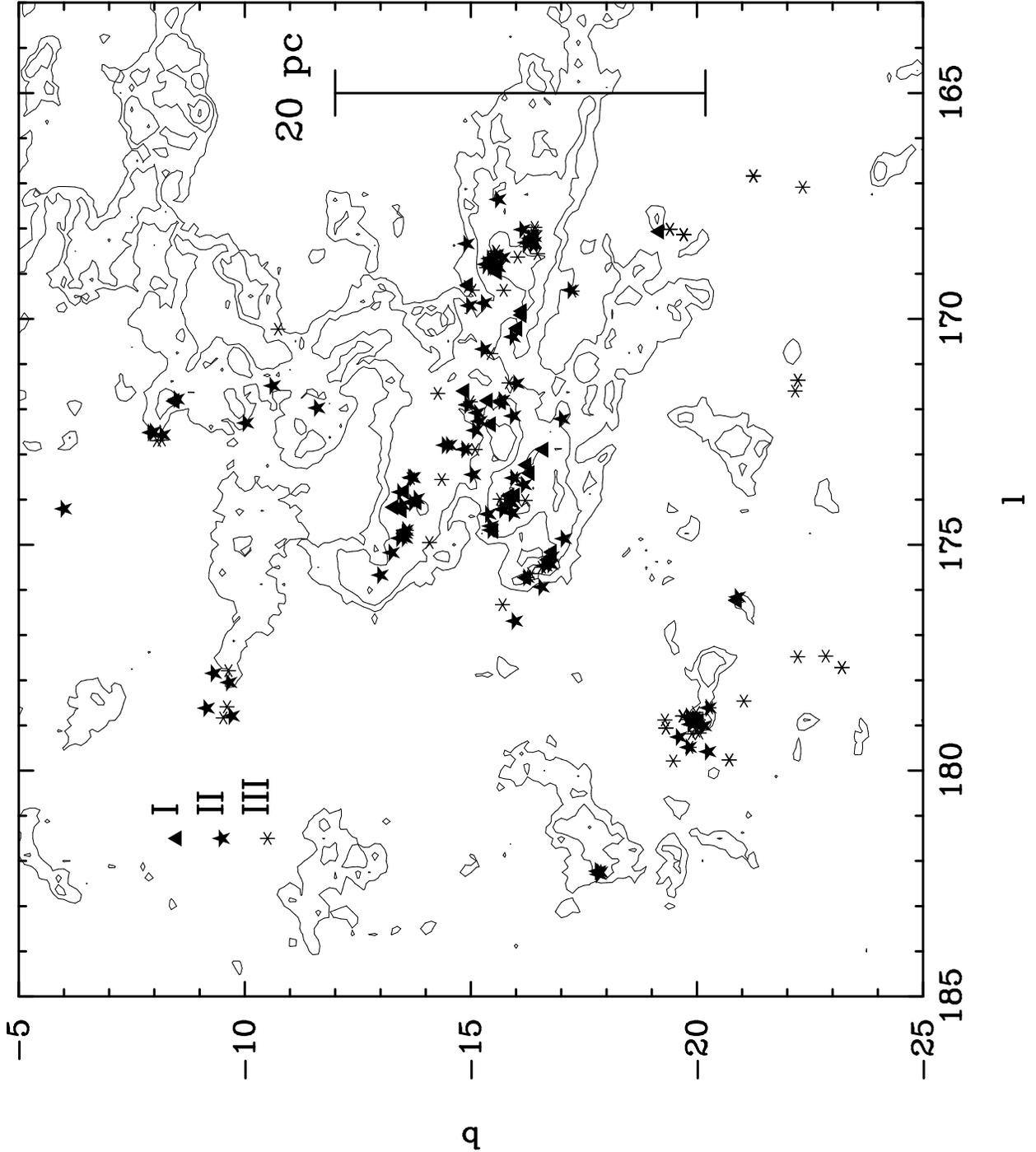}
\caption{Distribution of young stars in the Taurus molecular cloud,
superimposed upon $^{12}$CO emision contours taken from Megeath, Dame,
\& Thaddeus (2001; see Dame, Hartmann, \& Thaddeus 2001).  Class I,
II, and III sources are indicated, corresponding to protostars, stars
with disks, and stars without disks, respectively. The majority of the
stars in this plot have ages $\sim 2$~Myr, and an age spread not more
than $2-4$~Myr, even though the lateral extent of the region
approaches 40 pc and thus the associated crossing time is of order 20
Myr (see text).} 
\end{figure}

\begin{figure}[ht]
\plotone{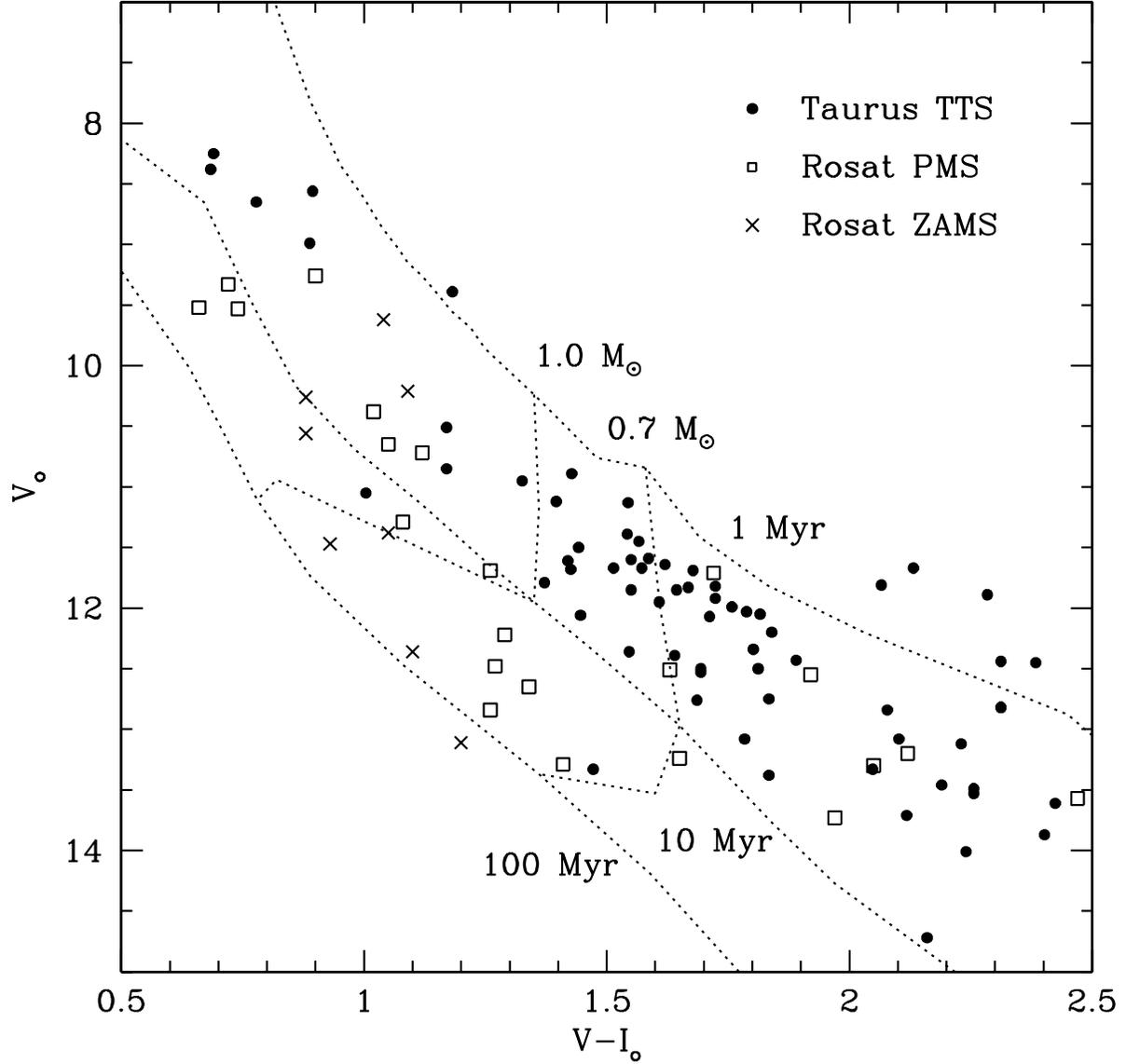}
\caption{Color-magnitude diagram comparing known young Taurus T Tauri
stars with ROSAT All-Sky survey sources.  The Taurus colors and
extinction corrections have been taken from Kenyon \& Hartmann (1995);
the ROSAT source data have been taken from Wichmann \etal (1999), as
discussed in the text.  Isochrones and tracks are from Siess \etal
(2000). The ROSAT sources are not numerous enough to represent a
significant star formation epoch in comparison with Taurus; moreover,
their substantially greater ages and wide spatial distribution imply
that they have mostly formed in widely-dispersed clouds which no
longer exist (see text).} 
\label{fig:HR}
\end{figure} 

\begin{figure}[ht]
\plotone{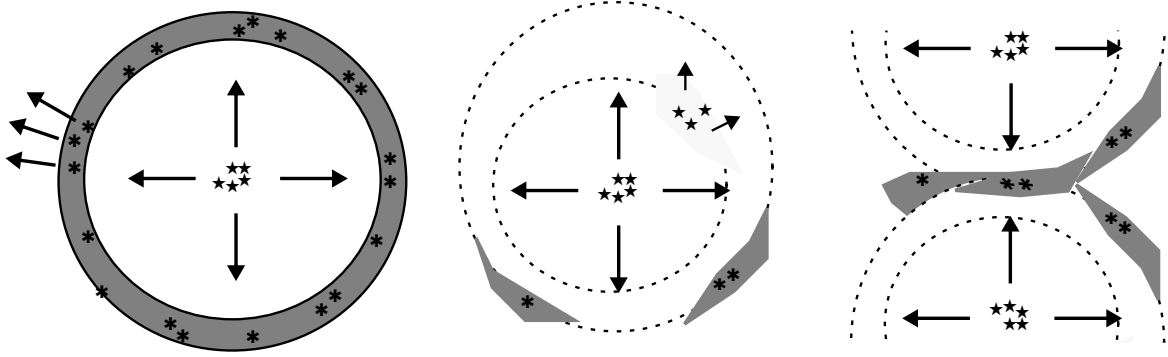}
\caption{A schematic view of large-scale triggered star formation.  In
the simplest possible case (left), star formation produces an
expanding shell which eventually becomes self-gravitating.  Densities
in the swept-up shell may therefore achieve similar values over very
large scales, and thus star formation can be coordinated over
timescales much shorter than the lateral crossing time (the bubble
radius divided by the velocity dispersion of the shell).  The
velocity dispersion among the stars formed in this gas can be small,
much less than the overall expansion velocity of the shell, over
regions small in comparison with the shell radius. Because the
interstellar medium is unlikely to be uniform on large scales, the
shell density will not be uniform, and so star formation cannot be
simultaneous over the entire shell (center), implying a range of star
formation epochs. However, if clouds are rapidly dispersed by the
star-forming event, the lifetime of molecular gas in any particular
region can still be short.  In general, the large-scale numerical
simulations of the ISM by PVP, BHV, etc. suggest that most clouds will
be formed from the interactions of flows from distinct star-forming
sites (right).} 
\end{figure}
 
\begin{figure}[ht]
\plotone{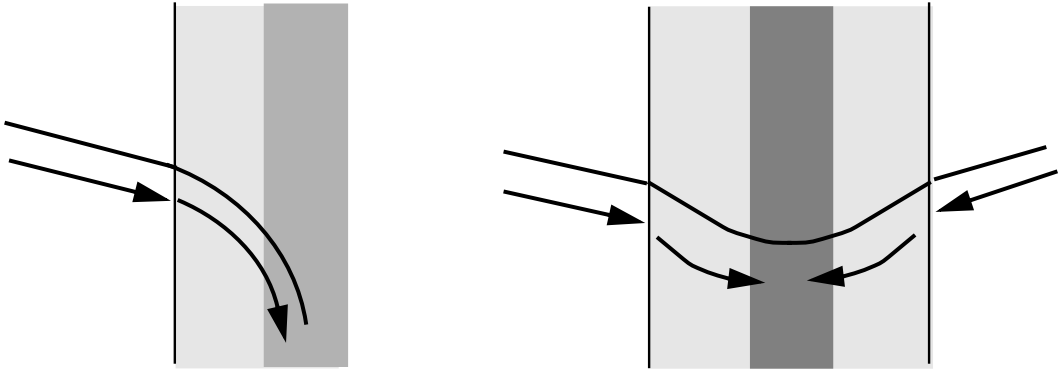}
\caption{Left panel: assumed field geometry for 1-dimensional steady
MHD oblique shock model. The field becomes increasingly tangential as
the post-shock gas cools, resulting in a magnetic pressure that
eventually limits compression. Right panel: typical geometry of clouds
seen in numerical simulations with magnetic fields in roughly
equipartition strength with the turbulent gas pressure. The clouds
tend to form in kinks or bends in the field lines, and dense regions
in the post-shock gas arise preferentially near where the tangential
magnetic field becomes small (see text).} 
\end{figure}

\begin{figure}[ht]
\plotone{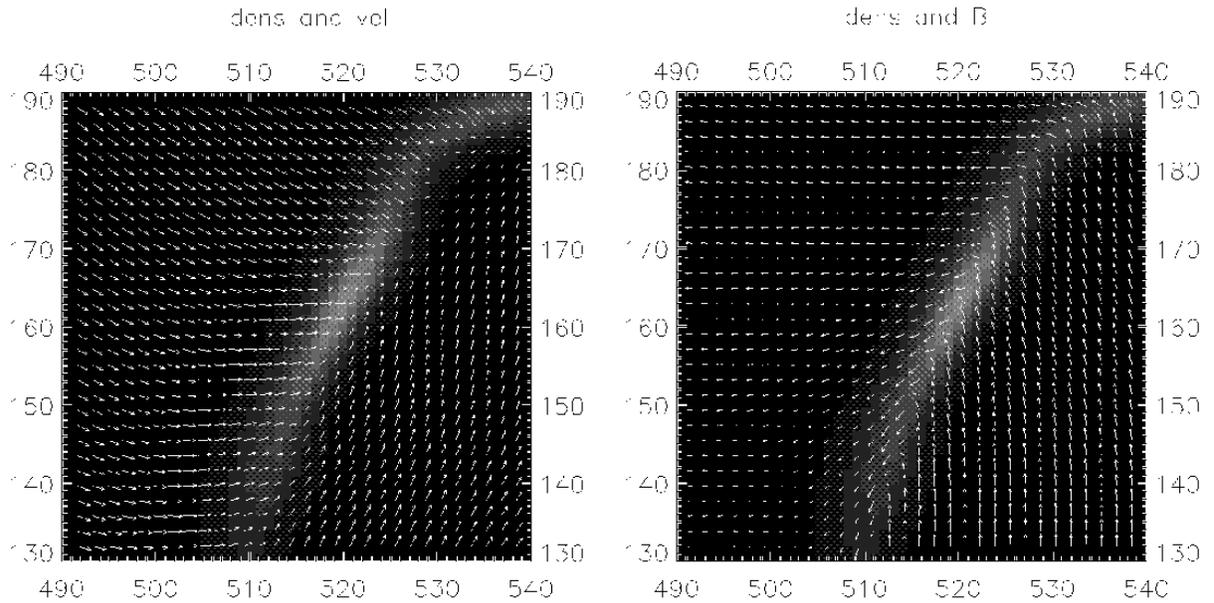}
\caption{Velocity (left) and magnetic (right) fields superimposed on
density grayscale levels for the cloud discussed in BHV.  The
formation of the cloud at a bend or ``kink'' in the magnetic field is
evident. 
\label{vbrho}}
\end{figure}

\begin{figure}[ht]
\plotone{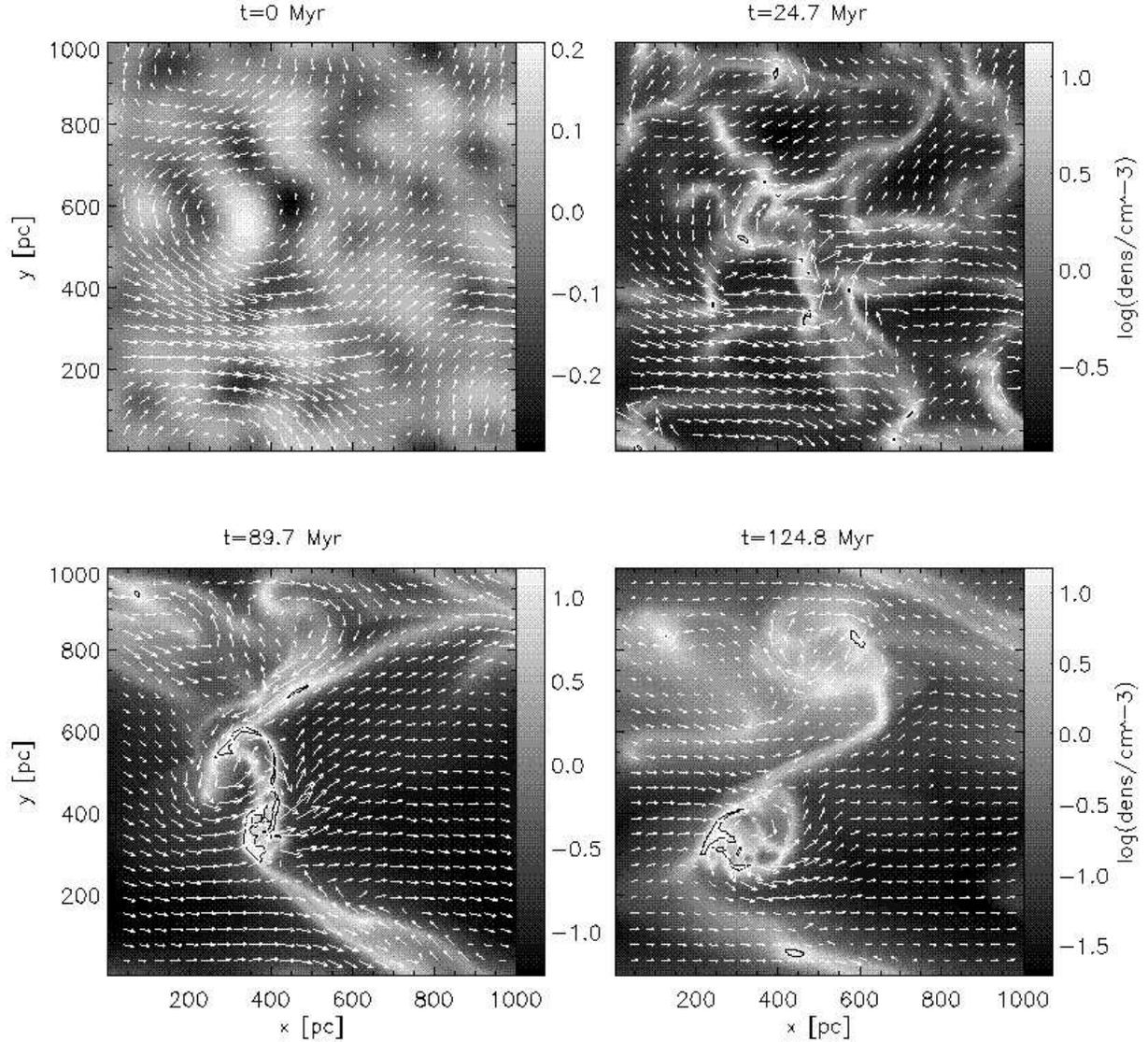}
\caption{Four snapshots of run cfa28 (see text) at (a) t=0, (b) t=24.7
Myr (c) t=89.7 Myr, and (d) t=124.8 Myr.  Vectors indicate magnetic
field directions and strengths. The grayscale denotes the density in
logarithmic units, as indicated in the grayscale bars.  ``Clouds'',
defined as regions where the density exceeds 15~cm\alamenos 3, are
denoted by the black isocontours. After about 10 Myr, ``star
formation'' occurs in the model (when local densities increase to the
threshhold level; see text), adding energy to the simulation.  Clouds
are built up by flows over scales of several hundreds of pc,
concentrating most of the mass into a small fraction of the
computational region. 
\label{acumula}}
\end{figure}

\begin{figure}[ht]
\plotone{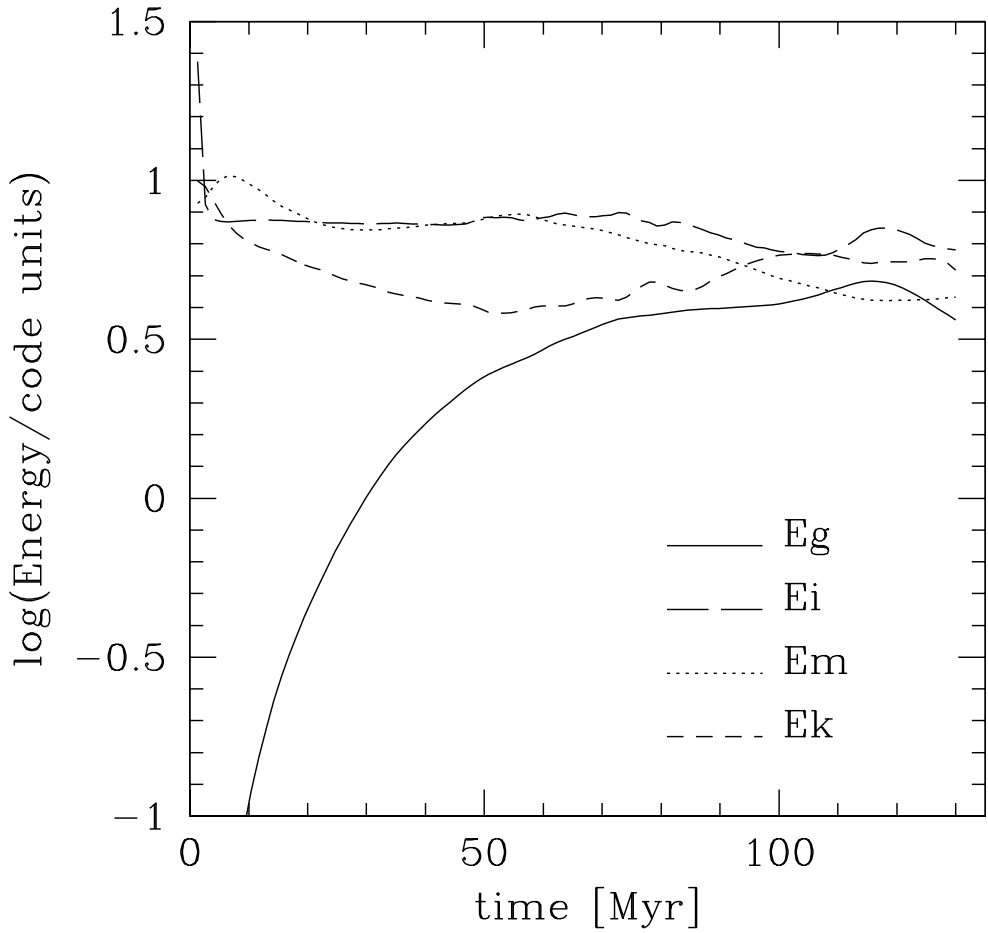}
\caption{Evolution of (negative) gravitational ($E_g$, solid line),
thermal ($E_i$, long-dashed line), magnetic ($E_m$, dotted line), and
kinetic ($E_k$, short-dashed line) energies for the whole
computational domain. Note that the internal energies are larger than
the gravitational energy, indicating that the whole computational
domain is supported against collapse. 
\label{fig:caja}}
\end{figure}

\begin{figure}[ht]
\plotone{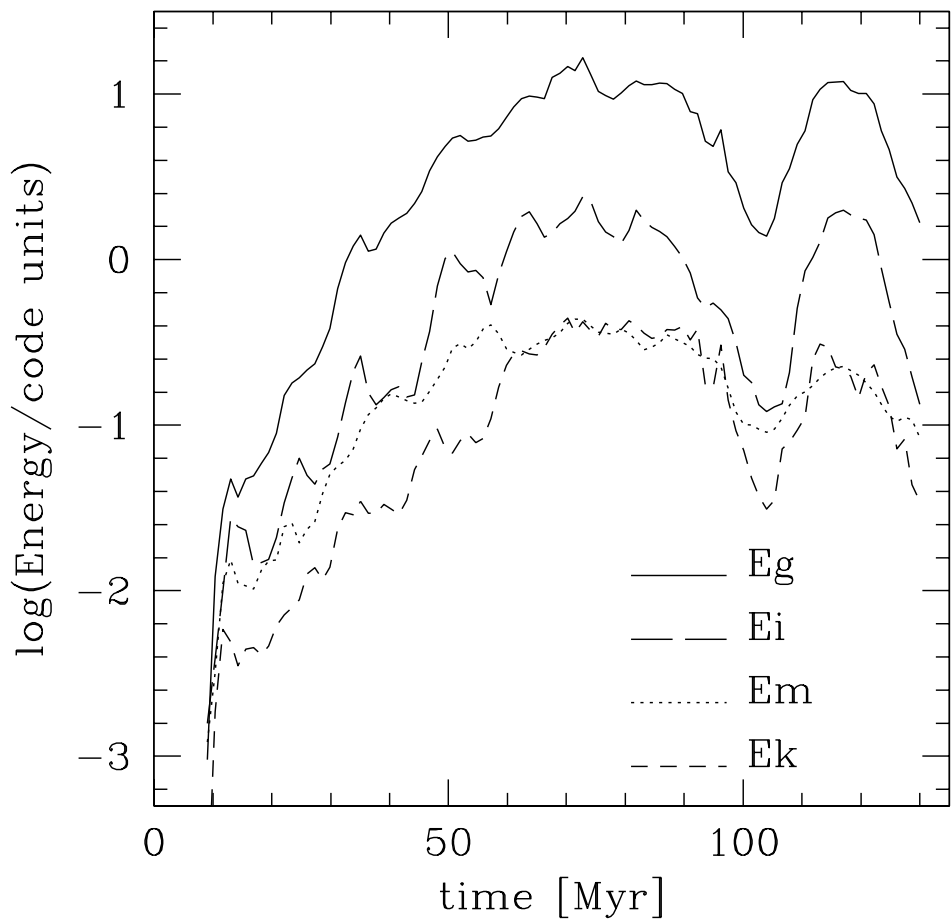}
\caption{ Similar to Fig.~\ref{fig:caja}, but for clouds (connected
set of pixels with densities above 15~cm\alamenos 3). Note that in
this case the larger energy is the gravitational, indicating that the
clouds are not supported against gravity, and can collapse
rapidly. The kinetic energy is calculated in the frame of reference of
the cloud.  
\label{fig:nube}}
\end{figure}

\begin{figure}[ht]
\plotone{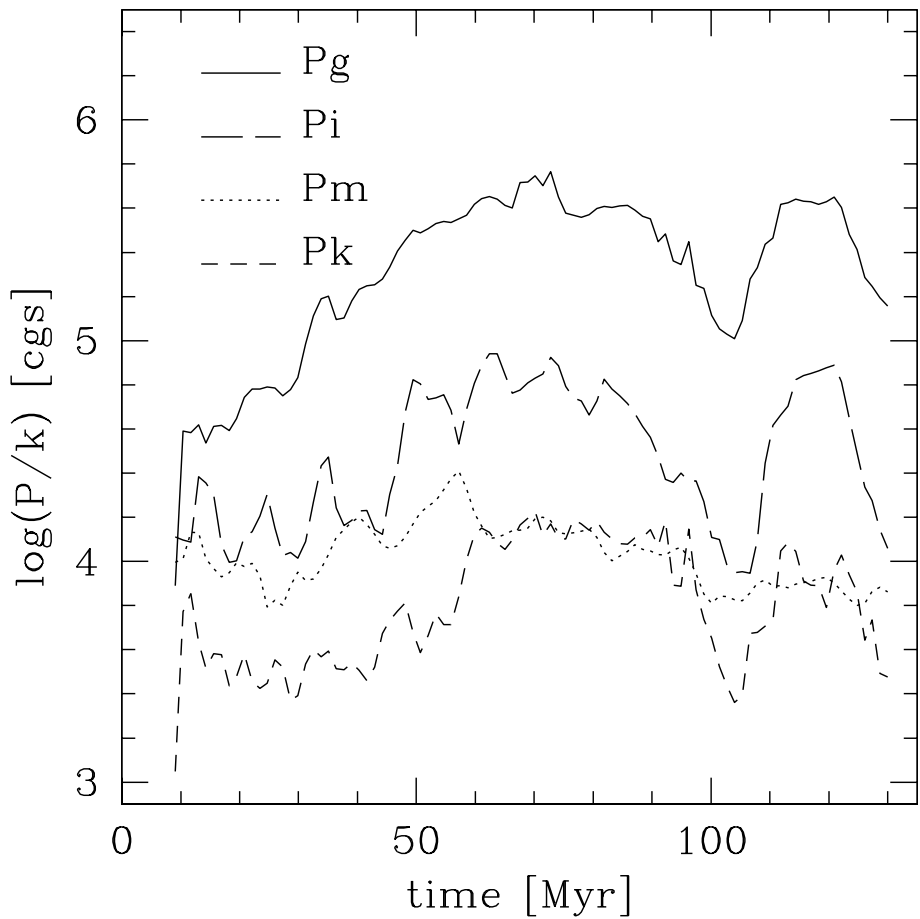}
\caption{Evolution of the pressures for the clouds in the
simulation. Each pressure is calculated as the energy divided by its
volume (in the present case, by its area, since the simulations are
two-dimensional), such that even the gravitational energy has its
counterpart in pressure. In this case, the gravitational pressure has
to be considered as a compressible pressure, since the effect of
gravity is the opposite to the magnetic or thermal pressure.  The
kinetic energy component in general includes the compressible
(contraction, expansion), and the incompressible (rotational)
components. Note that the values of the typical pressures (magnetic,
kinetic and thermal) are $P/k \sim 10^3 - 10^4 \cm^{-3}$~K. Comparison
with Figure \ref{fig:nube} indicates that much of the cloud energy
evolution is mostly due to changing cloud volumes (areas).
\label{fig:pvst}}
\end{figure}

\begin{figure}[ht]
\plotone{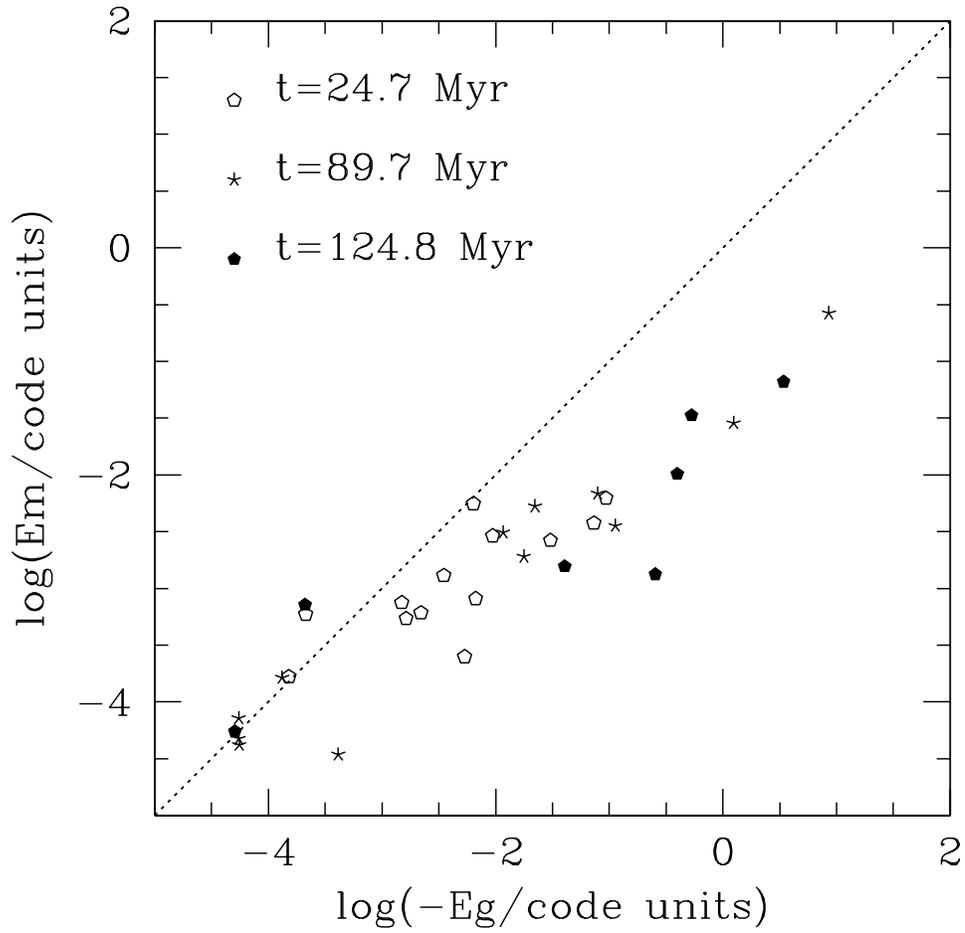}
\caption{Magnetic vs gravitational energy for clouds in the last 3
timesteps in Fig.\ref{acumula} ($t=$24.7, 89.7 and 124.8 Myr). Note
that while the energies are comparable, and correlated, the
gravitational energy is somewhat larger for more massive clouds,
implying that they are magnetically supercritical (see text). 
\label{fig:emvseg}}
\end{figure}

\clearpage

\begin{deluxetable}{c c c c}
\tablecolumns{4}
\tablecaption{Star forming regions \label{tbl-1}}
\tablehead{ \colhead{Region} & \colhead{$<t>$~(Myr)\tablenotemark{a}} 
& \colhead{Molecular gas?} &  \colhead{Ref. (age)} }
\startdata
Coalsack   & --    & yes &  -- \\
Orion Nebula&  1   & yes &  1  \\
Taurus      &  2   & yes &  1,2,3 \\
Oph         &  1   & yes &  1  \\
Cha I,II    &  2   & yes &  1  \\
Lupus       &  2   & yes &  1  \\   
MBM 12A     &  2   & yes &  10 \\
IC 348      &  1-3 & yes &  1,5,10 \\  
NGC 2264    &  3   & yes &  1  \\
Upper Sco   &  2-5   & no & 1,5,6  \\
Sco OB2     &  5-15 & no &  7  \\
TWA         &  $\sim$~10   & no &  8  \\
$\eta$~ Cha     & $\sim$~10   & no &  9  \\
\enddata

\tablenotetext{a}{Average age in Myr}

\tablerefs{(1) Palla \& Stahler 2000; (2) Hartmann 2001; (3) White \&
Ghez 2001; (4) Herbig 1998; (5) Preibisch \& Zinnecker 1999; (6)
Preibisch, G\"unther, \& Zinnecker 2001; (7) de Geus \etal 1989; (8)
Webb \etal 1999; (9) Mamajek, Lawson, \& Feigelson 1999; (10) Luhman
2001.} 
\end{deluxetable}

\end{document}